\documentclass[fleqn]{2023SCGE}
\usepackage{multirow}
\usepackage{color}

\newcommand{\hmpc}{\,h^{-1}\,{\rm Mpc}}

\newcommand{\hmsun}{\,h^{-1}\,{\rm M}_\odot}

\newcommand{\mpeak}{M_{\rm peak}}
\newcommand{\vpeak}{V_{\rm peak}}

\begin{document}

\ensubject{subject}

\ArticleType{Article}
\SpecialTopic{SPECIAL TOPIC: }
\Year{2026}
\Month{January}
\Vol{XX}
\No{X}
\DOI{??}
\ArtNo{000000}
\ReceiveDate{XXX}
\AcceptDate{XXX}

\title{What can galaxy clustering really tell us about the galaxy-halo connections?}

\author[1,2]{Xiaoju Xu}{{xiaojuxu@shnu.edu.cn}}%
\author[2,3]{Xiaohu Yang}{{xyang@sjtu.edu.cn}}
\author[2,3]{Zhongxu Zhai}{}
\author[2,3]{Yiyang Guo}{}%
\author[2,3]{Yizhou Gu}{}%
\author[2,3]{\\Yirong Wang}{}
\author[2,3]{Jiaxin Han}{}
\author[2,3]{Zhenlin Tan}{}
\author[2,3]{Junde Li}{}

\AuthorMark{Xu X}

\AuthorCitation{Xiaoju X, Yang X, et al}

\address[1]{Shanghai Key Lab for Astrophysics, Shanghai Normal University, \\Shanghai 200234, People’s Republic of China}
\address[2]{State Key Laboratory of Dark Matter Physics, Tsung-Dao Lee Institute \& School of Physics and Astronomy, \\Shanghai Jiao Tong University, Shanghai 201210, China}
\address[3]{Shanghai Key Laboratory for Particle Physics and Cosmology, and Key Laboratory for Particle Physics, Astrophysics and Cosmology, \\Ministry of Education, Shanghai Jiao Tong University, Shanghai 200240, China}


\abstract{Subhalo abundance matching (SHAM) is a commonly used framework for modeling the galaxy–halo connection. Yet, its standard implementation has difficulty reproducing the observed galaxy clustering with high accuracy (e.g., $\chi^2/{\rm dof}\sim1$). To overcome this issue, we propose a novel CS-SHAM framework, in which central and satellite galaxies are independently matched to main and satellite subhalos in simulations. Within this scheme, we introduce three free parameters to explicitly characterize the satellite fraction, $f_{\rm sat}$, as a function of stellar mass or absolute magnitude. 
To evaluate the performance of CS-SHAM, we apply it to two sets of mock galaxy catalogs built with the conventional SHAM method but using different subhalo mass proxies, $M_{\rm peak}$ and $V_{\rm peak}$, as well as two additional galaxy samples generated from a SAM and from TNG-300. We demonstrate that CS-SHAM reliably reproduces galaxy clustering whether $M_{\rm peak}$ or $V_{\rm peak}$ is used as the subhalo mass proxy. We also find that the models are unable to place robust constraints on $f_{\rm sat}$ if different mass proxies are employed. Indeed, within the CS-SHAM framework the halo occupation distribution (HOD) and conditional luminosity or stellar mass function (CLF/CSMF) are accurately recovered. Furthermore, we demonstrate for the first time that galaxy clustering constrains the HOD and CLF/CSMF primarily for relatively massive halos. Because the halo bias is nearly constant for low-mass halos, galaxy clustering is generally not very sensitive to the satellite population residing in these low-mass systems. }


\maketitle


\begin{multicols}{2}
\section{Introduction}
\label{sec:Intro}

Within the paradigm in which galaxies form and evolve inside the gravitational potential wells of dark matter halos \cite{White1978}, high-resolution N-body simulations \cite{Springel2005, Prada2012} have been instrumental in developing empirical models that connect galaxies with their host halos. These empirical techniques are essential for modeling galaxy clustering in wide-area spectroscopic surveys, such as the Sloan Digital Sky Survey (SDSS; \cite{york2000}), the Extended Baryon Oscillation Spectroscopic Survey (eBOSS; \cite{Dawson2016}), and the Dark Energy Spectroscopic Instrument (DESI; \cite{desi2016}). 
\Authorfootnote 

The Halo Occupation Distribution (HOD) and Conditional Luminosity Function (CLF) formalisms (e.g., \cite{Jing1998, Berlind02,Zheng2005,Zheng07, Yang03, Bosch03, Yang08, Yang2009})—which describe the abundance and luminosity (or stellar mass) distribution of galaxies of a given type as a function of halo mass—are widely used to interpret observed galaxy clustering \cite{Zehavi2005,Zehavi2011} and to constrain cosmological parameters \cite{Cacciato2013, More2015,Lange2019,Zhai2023,Zhai2024,Wang2025}. HOD models employ distinct halos identified in N-body simulations and typically assume that central galaxies occupy halo centers, while satellite galaxies trace an NFW profile \cite{Navarro1996}. The standard HOD framework characterizes volume-limited galaxy samples, defined by fixed thresholds in absolute magnitude or stellar mass, using five free parameters. Additional parameters are usually needed for more complex tracers, such as emission-line galaxies and luminous red galaxies \cite{Geach2012,Alam2020,Avila2020,Paviot2024}. Given that galaxy assembly bias (GAB)—the dependence of galaxy properties on secondary halo characteristics—has been shown to affect galaxy clustering \cite{Croton2007,Zehavi2018,Contreras2019,Hadzhiyska2020a,Xu2021a}, more recent model extensions incorporate GAB via extra parameters \cite{McEwen2018,yuan2020,Xu2021a,Salcedo2022,Zhai2023}.

While the HOD/CLF framework can reproduce observed galaxy clustering with high precision, it relies on a relatively large set of parameters, many of which must be re-tuned for each stellar-mass threshold, particularly in the HOD case. By contrast, Subhalo Abundance Matching (SHAM, e.g., \cite{Kravtsov2004,Conroy2006,Vale2006,Reddick2013}) links galaxy luminosity or stellar mass directly to subhalo mass proxies and characterizes the galaxy–halo connection over a wider luminosity/stellar-mass range. SHAM approaches make use of the subhalo population by matching the cumulative galaxy luminosity/stellar-mass function to the cumulative subhalo mass function, thereby assigning every galaxy to the center of a subhalo. Relative to the standard five-parameter HOD model, a basic SHAM implementation typically needs only a single free parameter—the scatter in galaxy luminosity/stellar mass at fixed subhalo mass—together with an assumed monotonic mapping between the two.
Apart from implementations that adopt an observed luminosity function, another family of SHAM models explicitly parameterizes the stellar-to-halo mass relation (SHMR) and constrains it using both the observed stellar mass function and galaxy clustering, thus introducing more parameters than the standard SHAM setup \cite{Behroozi2010,Yang2012,Behroozi2013,Moster2010,Moster2013}. As these works primarily aim to study galaxy formation, the resulting models are not particularly convenient for cosmological applications.

Despite the simplicity of basic SHAM, which involves only a single free parameter, it faces challenges in simultaneously matching the observed galaxy clustering on both small and large scales \cite{Guo2016,Campbell2018}. On small scales in particular, SHAM tends to systematically underpredict galaxy clustering. This tension can be alleviated by adopting subhalo mass proxies such as $V_{\rm peak}$—the maximum circular velocity reached over a subhalo’s merger history—instead of the present-day subhalo mass, because $V_{\rm peak}$ carries information about the formation history of the subhalo. Models based on $V_{\rm peak}$ have been shown to increase the satellite fraction within SHAM \cite{Reddick2013,Chaves-Montero2016,Rodriguez-Torres2016,Campbell2018}. Other alternative mass indicators have also been widely investigated \cite{Stiskalek2021,Tonnesen2021,DeRose2022,Chuang2023,Masaki2023}. 
Beyond changing the mass indicator, SHAM implementations often incorporate orphan galaxies—systems whose subhalos have been tidally disrupted after accretion onto a host halo but whose galaxies persist \cite{DeLucia2007,Moster2010,Guo2011,Guo2014}. Including orphans introduces extra parameters that increase the model’s flexibility on small scales \cite{Moster2013,Contreras2021b,DeRose2022}. In analogy with HOD approaches, further extensions have been proposed to incorporate GAB within SHAM in order to better reproduce galaxy clustering on large scales \cite{Lehmann2017,Contreras2021b,Contreras2021a,DeRose2022}.

However, these trial-and-error approaches are not flexible enough to reproduce the observed galaxy clustering and thus have limited use for cosmological applications, with the exception of SHAM models that include orphan galaxies, which have been shown to mitigate this issue and enable constraints on cosmological parameters\cite{Contreras2023,Mahony2026}. In this study, we present a novel approach, CS-SHAM, which carries out abundance matching for central and satellite galaxies separately. In line with the fundamental idea of HOD/CLF models, the strength of galaxy clustering is highly sensitive to the abundance of satellite galaxies. In the limiting cases, $f_{\rm sat}=1$ leads to the strongest clustering signal, whereas $f_{\rm sat}=0$ results in the weakest one \cite{Yang2012}. We therefore introduce an $f_{\rm sat}$ parameterization to achieve the flexibility required to match the observed galaxy clustering. Indeed, similar treatments of the satellite fraction have been adopted in several recent studies of ELG galaxy clustering \cite{Favole2016, Gao2022, Gao2023, Yu2024}, employing either a constant or halo-mass-dependent $f_{\rm sat}$.

Since galaxy clustering is typically measured in bins of {\it observed} luminosity or stellar mass, we explicitly parametrize the satellite fraction as a function of galaxy stellar mass or absolute magnitude, thereby capturing how clustering depends on stellar mass or luminosity. We first assess our model using mock galaxy catalogs constructed with a basic SHAM framework, in which the sole free parameter is the scatter in the stellar-to-halo mass relation \cite{Gu2024}. We then further test the model with semi-analytic galaxy catalogs (SAM; \cite{Guo2011,Guo2013,Luo2016}) and with hydrodynamical simulations \cite{Schaye2015,Nelson2018,Nelson2019}. For all galaxy samples, we use $M_{\rm peak}$ and $V_{\rm peak}$ as proxies for subhalo mass and compare how well each performs in reproducing the observed galaxy clustering and satellite fractions.

The remainder of this paper is organized as follows. In Section \ref{sec:Data}, we describe the mock galaxy catalog, the SAM, and the hydrodynamic simulation employed in our analysis. Section \ref{sec:sham_emu} introduces our CS-SHAM framework, along with the associated measurements and emulator. In Section \ref{sec:results}, we present the outcomes of applying our fiducial CS-SHAM model to the mock galaxy catalogs, the SAM, and the hydrodynamic simulation. Section \ref{sec:HOD-CLF} examines the HOD and CLF/CSMF that our CS-SHAM approach is able to reproduce. Finally, Section \ref{sec:summary} provides a summary of our findings and a discussion of their implications.

\section{Simulations and galaxy catalogs}
\label{sec:Data}
In this section, we briefly introduce the galaxy samples we used to evaluate the performance of our CS-SHAM model, including mock galaxies produced by basic SHAM, semi-analytic galaxies, and hydrodynamic simulation.

\subsection{Jiutian simulation and CSST mock galaxy catalog}
\label{sec:mock}

We first validate our CS-SHAM using mock galaxy catalogs constructed from Jiutian simulation \cite{Han2025}. As a high-resolution N-body simulation suite designed for the China Space Station Telescope (CSST; \cite{Gong2019}), Jiutian consists of four modules: the primary runs with high resolutions for fiducial Planck 2018 cosmology with $\Omega_{\rm m}=0.3111$, $\Omega_{\rm b}=0.049$, and $n_s = 0.9665$ and $\sigma_8=0.8102$ \cite{Planck2020}, the emulator runs exploring the parameters around the fiducial cosmology \cite{Chen2025a}, the reconstruction runs recovering the observed large-scale structure, and the extension runs employing extended cosmologies beyond the standard model. 
The primary runs are carried out in boxes of 0.3, 1 and 2 $h^{-1}{\rm Gpc}$, evolving $6144^3$ particles of mass $1.005 \times 10^7 \hmsun$, $3.723 \times 10^8 \hmsun$, and $2.978 \times 10^9 \hmsun$, respectively. The 0.3 and 2 $h^{-1}{\rm Gpc}$ runs adopt the GADGET-4 code \cite{Springel2021} and the 1 $h^{-1}{\rm Gpc}$ run adopts LGADGET-3 code \cite{Angulo2012}. Halos are identified by the FOF algorithm \cite{Davis1985}, and subhalos are identified by the Hierarchical Bound-Tracing method (HBT+, \cite{Han2012,Han2018}). The HBT+ algorithm offers the advantage of generating consistent subhalo catalogs, naturally tracing disrupted subhalos and decomposing subhalos according to merger levels. 

Mock galaxy catalogs are produced based on the subhalo lightcone of the Jiutian 1 $h^{-1}{\rm Gpc}$ box using basic SHAM \cite{Gu2024}. The mocks consist of three catalogs constructed using $M_{\rm peak}$ (peak mass along the merger history of subhalo), $V_{\rm max}$ (maximum circular velocity of subhalo), and $V_{\rm peak}$ (peak value of maximum circular velocity of subhalo along the merger history) as mass or luminosity indicators, respectively. We make use of the $M_{\rm peak}$ and the $V_{\rm peak}$ mocks. Galaxy $z$-band luminosities are assigned to subhalos by matching the cumulative galaxy luminosity function measured from DESI Y1 \cite{Wang2024} to the cumulative subhalo mass or velocity functions of the above indicators. The only parameter, $\sigma_{{\rm log}L}$, which quantifies the scatter between $z$-band luminosity and subhalo mass indicators is set to be 0.15. The mock catalogs covers a sky area of $\sim18200$ ${\rm deg}^2$ and spans a redshift range of $z=[0,1]$. For simplicity, we restrict our analysis to galaxies within the redshift range $z=[0,0.3]$.

We emphasize that the subhalo lightcone used in the mock catalogs is generated from simulation snapshots, with subhalo positions and properties obtained via interpolation. To maintain consistency in testing our CS-SHAM method, we use the same algorithm to build the subhalo lightcone catalog, where subhalo trajectories are interpolated between snapshots \cite{Tan2026}. Readers seeking more information on the algorithms used to populate subhalos with mock galaxies, as well as on duplication of simulation boxes and the implementation of observational selection effects, etc., are referred to \cite{Gu2024} for further details.

\subsection{SAM and hydrodynamic galaxies}
\label{sec:SAMhydro}

To assess the performance of CS-SHAM, we also utilize galaxies from a SAM \cite{Luo2016} implemented on the ELUCID $N$-body simulation \cite{Wang2016}. ELUCID is a constrained simulation designed to reproduce the large-scale structure of SDSS DR7 \cite{Abazajian2009}. It employs a reconstructed matter density field derived from the group catalog \cite{Wang2012} and an initial density field inferred by the Hamiltonian Monte Carlo Markov chain method \cite{Wang2014}. Adopting WMAP5 cosmology with cosmological parameters $\Omega_{\rm m}$ = 0.258, $\Omega_{b}$ = 0.044, $h$ = 0.72, and $n_s$ = 0.963, and $\sigma_8$ = 0.796 \cite{Dunkley2009}, ELUCID evolves $3072^3$ particles of mass $3.0875 \times 10^8 \hmsun$ 
within a box of 500$h^{-1}$Mpc using the GADGET-2 code \cite{Springel2005}. The SAM in \cite{Luo2016} is a modified version of the L-Galaxies model \cite{Guo2011,Guo2013} improving the trace of low-mass subhaloes for modeling satellite quenching and galaxy clustering.

We further incorporate galaxies from the TNG-300 hydrodynamic simulation \cite{Marinacci2018, Naiman2018, Nelson2018, Nelson2019, Pillepich2018, Springel2018}. Using the \texttt{AREPO} code \cite{Springel2010}, TNG-300 evolves $2500^3$ dark matter particles of mass $5.9 \times 10^7 \hmsun$ and $2500^3$ baryonic particles of mass $1.1 \times 10^7 \hmsun$ in a box of length 205 $h^{-1}{\rm Mpc}$. The simulation adopts Planck cosmology with $\Omega_{\rm m}=0.31$, $\Omega_{\rm b}=0.0486$, $h=0.677$, and $n_s$=0.97, and $\sigma_8=0.816$ \cite{Planck2016}.

To minimize the impact of cosmology, simulation volume, and resolution on our model constraints, we fit the CS-SHAM model by matching the galaxy clustering in the ELUCID SAM and TNG-300 to that of subhalos drawn from the corresponding ELUCID N-body run and the dark-matter-only TNG-300-dark simulation, respectively. In the CS-SHAM framework, we adopt $M_{\rm peak}$ and $V_{\rm peak}$ as subhalo mass proxies to reproduce the galaxy clustering in both the SAM and the TNG-300 simulations. All simulations and galaxy samples used in this work are listed in Table \ref{tab:simulation}.

\begin{table*}
\centering
\footnotesize
\caption{Galaxy samples and the corresponding simulations used. }
\label{tab:simulation}
\tabcolsep 18pt
\begin{tabular*}{\textwidth}{lccc}
\toprule
galaxy sample name & $N$-body simulation & model or code of sample producing & CS-SHAM model applied\\ 
\hline
Jiutian catalogs & Jiutian (lightcone) & basic one-parameter SHAM & fiducial four-parameter \\
\hline
ELUCID SAM  & ELUCID simulation & L-Galaxies & fiducial four-parameter \\
\hline
TNG-300 & TNG-300 hydrodynamic simulation  & \texttt{AREPO}  & fiducial four-parameter \\
\bottomrule
\end{tabular*}
\end{table*}


\section{CS-SHAM model and emulator} 
\label{sec:sham_emu}

\subsection{CS-SHAM with satellite fraction}
\label{sec:sham}

\begin{figure*}[!ht]
\centering
\includegraphics[width=0.85\textwidth]{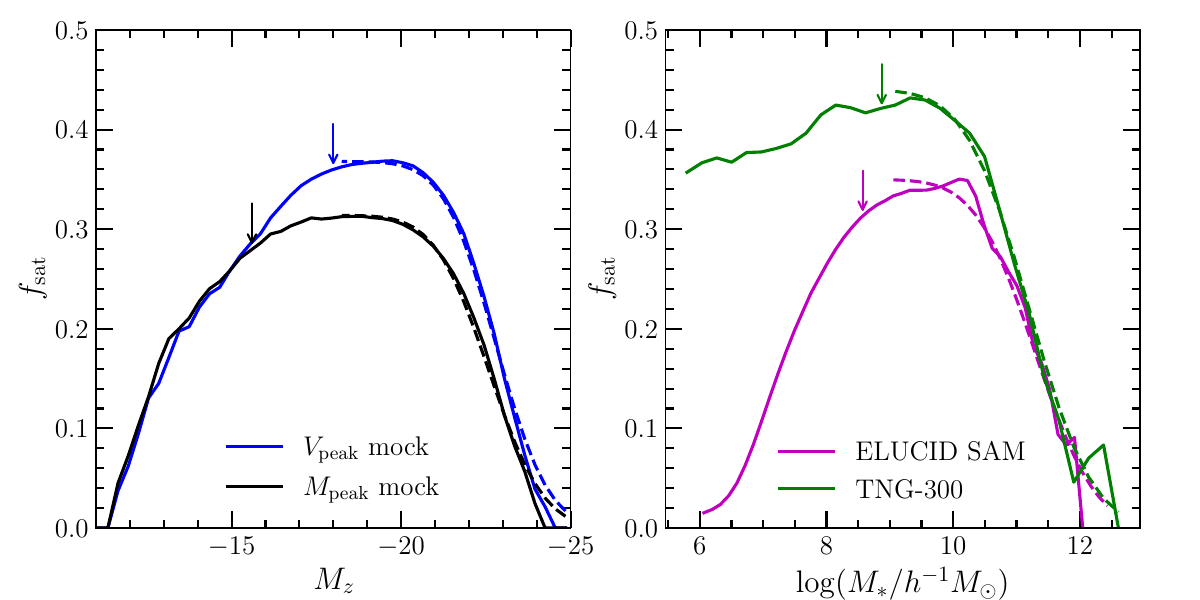}
\caption{Satellite fraction $f_{\rm sat}$ measured in the Jiutian mocks, ELUCID SAM, and TNG-300. The left panel presents $f_{\rm sat}$ as a function of z-band absolute magnitude for Jiutian mocks constructed using $M_{\rm peak}$ (black solid) and $V_{\rm peak}$ (blue solid). The right panel displays $f_{\rm sat}$ as a function of stellar mass for ELUCID SAM (magenta) and TNG-300 (green). In both panels, the dashed curves show the best-fitting models to $f_{\rm sat}$, obtained with Equation \ref{eq:fsat1} for the Jiutian mocks over $M_{z}>-18$, and with Equation \ref{eq:fsat2} for ELUCID SAM and TNG-300 over ${\rm log} M_{*}>9$. Resolution limits of each galaxy sample are indicated by arrows.}
\label{fsat}
\end{figure*}

To retain flexibility in modeling the clustering strength of galaxies, CS-SHAM assigns central and satellite galaxies independently to central and satellite subhalos, and uses extra free parameters to define the satellite galaxy fraction as a function of stellar mass or luminosity. With this strategy, there is no need to introduce orphan galaxies to adjust small-scale clustering, and the method is generally insensitive to the particular choice of subhalo mass indicator. Before parameterizing the satellite fraction $f_{\rm sat}$, we first measure it from all the galaxy samples we used (i.e., the Jiutian catalogs, ELUCID SAM, and TNG-300) and show the results in Figure \ref{fsat}. The left panel presents $f_{\rm sat}$ as functions of $z$-band absolute magnitude, measured from the standard Jiutian mocks in Section \ref{sec:mock} built based on $M_{\rm peak}$ and $V_{\rm peak}$ in solid colored curves. $f_{\rm sat}$ increases from bright end, peaks at $M_{z}\sim-18$, and declines for fainter galaxies. This behavior aligns with the result of \cite{Wang2024}, where the satellite fraction is inferred from a group catalog \cite{Yang2021} based on both photomatric redshift and spectropic redshift from DESI Legacy Surveys DR9. In this work, we focus only on the bright regime ($M_{z}<-18$) and parameterize $f_{\rm sat}$ in this range using an error function: 
\begin{equation} \label{eq:fsat1}
f_{\rm sat}(M_z)=f_1\cdot{\rm erf}[(M_z-f_2)\cdot f_3]+f_1
\end{equation}
where $f_1$ marks half the amplitude of $f_{\rm sat}$, and $f_2$ is the characteristic magnitude where $f_{\rm sat}(f_2)=f_1$, and $f_3$ quantifies the width of $f_{\rm sat}$ compared to an standard error function. The colored dashed curves indicate the best-fit to the mock $f_{\rm sat}$ using this parameterization. To ensure the robustness and completeness of the galaxy samples, we adopt resolution limits of $M_z$ (indicated by colored arrows) for the two mocks. These limits are determined by the upper bounds of $M_z$ in the $M_z-M_{\rm peak}$ relation, where $M_{\rm peak}$ corresponds to 100 dark matter particles. 

In the right panel, we show $f_{\rm sat}$ as a function of stellar mass for both the ELUCID SAM and TNG-300. As in the mock catalogs, $f_{\rm sat}$ in these samples increases toward lower stellar masses and then drops below ${\rm log}M_*\sim 9$. Similar to the mocks, the resolution limit of $M_*$ in ELUCID SAM is determined with the $M_*-M_{\rm sub}$ relation where the subhalo contains 100 dark matter particles. For TNG-300, we adopt a limit of $M_*$ corresponding to 100 stellar mass particles.
In this paper, for galaxies with ${\rm log}M_{*}>9$, we use this function,
\begin{equation} \label{eq:fsat2}
f_{\rm sat}(M_*)=f_1\cdot{\rm erf}[(f_2-M_*)\cdot f_3]+f_1.
\end{equation}
to describe the satellite fraction. The dashed curves shown in the right panel of Fig. \ref{fsat} represent the best fits obtained from Equation \ref{eq:fsat2} for galaxies with ${\rm log}M_*  > 9$.
  
Based on these $f_{\rm sat}$ parameterizations, the CS-SHAM model assigns $M_z$ or stellar masses to subhalos through the following steps:
\begin{itemize}
    \item[(1)] For each subhalo in Jiutian lightcone catalog, we assign an $M_z$ value to it matching the cumulative subhalo mass function to cumulative $M_z$ function. The $M_z$ function adopted is measured from DESI Y1 in \cite{Wang2024}, which is the same as that used to build the mock catalog in \cite{Gu2024}. In this step, an overall $M_z$ list is produced for all subhalos. For subhalos in ELUCID and TNG-300, stellar masses are assigned by matching the respective cumulative subhalo mass functions to the stellar mass functions measured directly within each galaxy catalog, generating corresponding stellar mass lists;
    \item[(2)] For each $M_z$ (or stellar mass) in the lists, we then calculate its probability of being satellite using Equation \ref{eq:fsat1} (or Equation \ref{eq:fsat2}) for fixed $f_{\rm sat}$ parameters, and then randomly classify it as central or satellite. This partitions the $M_z$ (or stellar mass) lists into distinct central and satellite lists;
    \item[(3)]The subhalo mass indicators and $M_z$ (or stellar mass) are each sorted independently in descending order, after which central and satellite galaxies are assigned to central and satellite subhalos, respectively, according to their ranks;
    \item[(4)] For centrals and satellites, scatter $\sigma$ in stellar-to-subhalo mass relation is added to $M_z$ (or stellar mass) values. However, this will tilt the  $M_z$ or stellar mass function, suppressing the high-mass end while enhancing the low-mass end; 
    \item[(5)] To resolve the issue above, we rank the $M_z$ (or stellar mass) again after including scatters. Then $M_z$ (or stellar mass) are matched to subhalos using the new rank but with the original values corresponding to the same rank. This maintains the original stellar mass function while incorporating scatters.
\end{itemize}

Previous studies adopt deconvolution methods to incorporate scatter in stellar mass at fixed subhalo mass in SHAM to recover the observed stellar mass function \cite{Behroozi2010}. Our approach provides a computationally efficient alternative that directly recovers both the target scatter and the stellar mass function
\cite{McCullagh2017,Gaines2021,Berti2023,Gu2024}. Other approaches such as introducing scatter to subhalo mass indicators rather than stellar mass are also adopted in SHAM \cite{Nuza2013,Yu2022,Masaki2023}.

\begin{figure*}[!htb]
\centering
\includegraphics[width=0.85\textwidth]{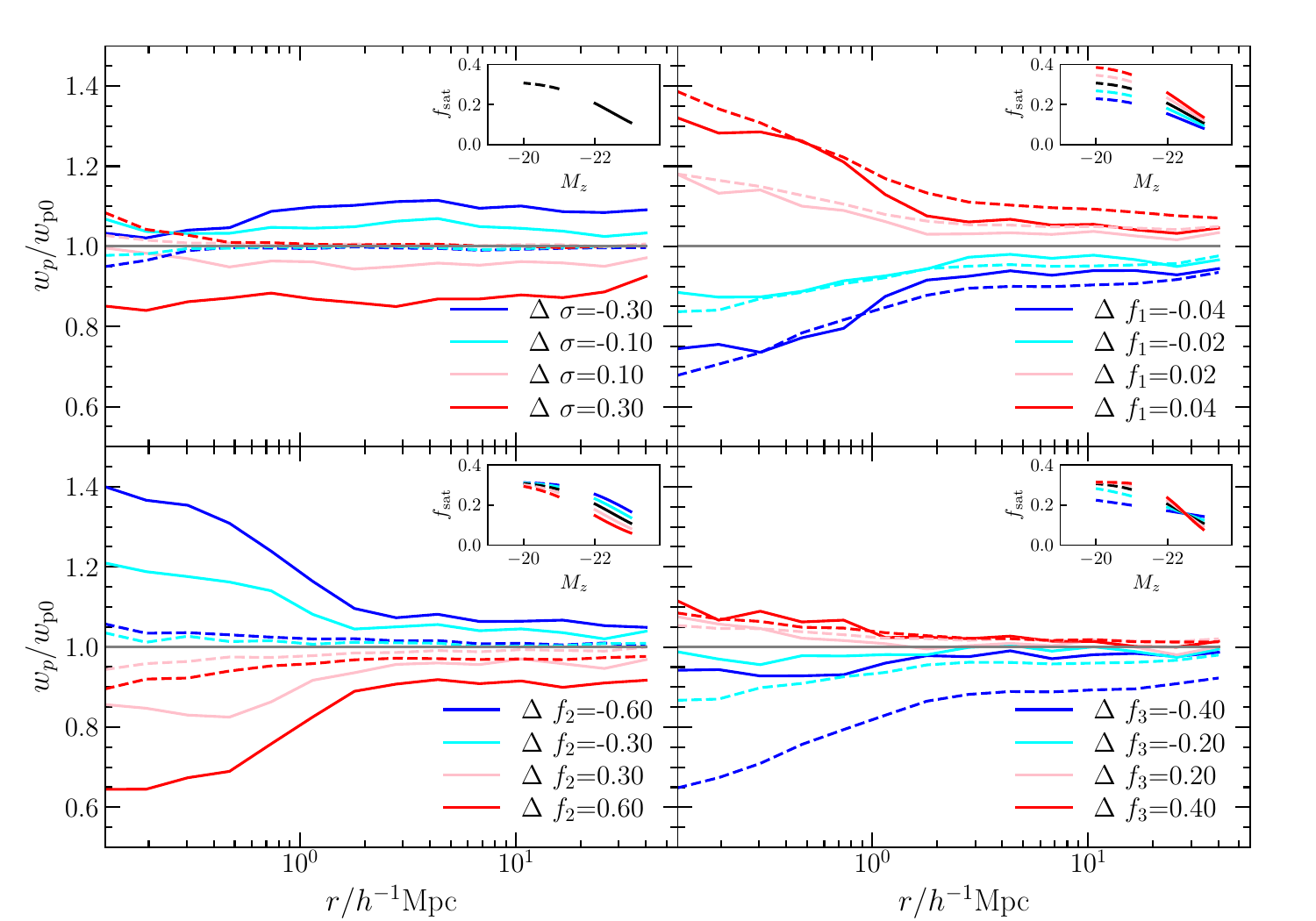}
\caption{Effects of CS-SHAM parameters on $w_p$ for the Jiutian subhalo lightcone catalog. The y-axis shows the ratio of $w_p$ to the fiducial clustering $w_{\rm p0}$, where $w_{\rm p0}$ is computed using the true parameter values from the mock in \cite{Gu2024}. The top-left, top-right, bottom-left, and bottom-right panels display the impact of varying $\sigma$, $f_1$, $f_2$, and $f_3$ around their true values, respectively. Red and pink curves represent increases in the parameter values, while blue and cyan curves represent decreases. Solid lines show the clustering ratios for $M_z=[-23,-22]$, and dashed lines how the ratios for $M_z=[-21,-20]$. Inset small panels show the corresponding $f_{\rm sat}$ for the varying parameters in each panel for both the two $M_z$ bins.}
\label{parameffect}
\end{figure*}

Using the procedure described above, we populate subhalos with galaxies and construct catalogs that contain $M_z$ (or stellar mass) and spatial coordinates for specified sets of model parameters. Within our CS-SHAM framework, there are in total only four free parameters: $\sigma$, $f_1$, $f_2$, and $f_3$. 
Before applying this to the mock galaxy catalogs introduced in Section \ref{sec:Data}, we first examine how variations in these four free parameters impact galaxy clustering, e.g., the projected two point correlation function, $w_p$. 
In this study, we use the Landy-Szalay estimator \cite{LandySzalay1993} to measure $w_p$ in multiple magnitude or stellar mass bins,
\begin{equation} \label{eq:wp}
w_p(r_p)=2\int_{0}^{r_{\rm max}} \xi(r_p,\pi) d\pi.
\end{equation}
For the Jiutian mock catalogs, we measure $w_p$ in 14 $r_p$ bins ranging from 0.1 to 50 $\hmpc$, with $r_{\pi \rm max}=100\hmpc$ for five $M_z$ bins from $M_z=-23$ to $M_z=-18$ for the mock fitting. For the ELUCID SAM (or TNG-300), we measure $w_p$ in 13 $r_p$ bins with $r_{\pi \rm max}=50\hmpc$ (or $r_{\pi \rm max}=30\hmpc$) for six stellar mass threshold bins from ${\rm log}M_*>9$ to ${\rm log}M_*>11.3$ (or ${\rm log}M_*>11.2$). The $w_p$ calculation in this work are performed using the {\tt CORRFUNC} package \cite{Sinha2020}. 

In Figure \ref{parameffect}, we present how the model parameters influence the projected correlation function $w_p$ in the CS-SHAM framework based on $M_{\rm peak}$, using the Jiutian subhalo lightcone catalog. The parameter $\sigma$, which characterizes the scatter between $M_z$ and $M_{\rm peak}$ at fixed ${\rm log}M_{\rm peak}$, affects $w_p$ on all scales. For $M_z$ = [-23, -22], increasing the scatter lowers the clustering amplitude, while for $M_z$ = [-21, -20], the impact of $\sigma$ is minimal. In general, the influence of $\sigma$ is small relative to that of the other parameters. In the $M_{\rm peak}$ mock, the fiducial value is $\sigma = 0.375$, and varying $\sigma$ by 0.1 changes $w_p$ by only $\sim 5\%$.

The parameter $f_1$, which sets the half-height point of the $f_{\rm sat}$ curve, strongly influences small-scale clustering in both $M_z$ bins. In the mock, the true value is $f_1=0.157$, and changing it by 0.02 leads to $\sim$10–20\% variations in $w_p$ at scales below 1 $\hmpc$. Increasing $f_1$ (i.e., raising the satellite fraction) boosts small-scale clustering by populating halos with more satellite galaxies, while its effect on larger scales is relatively modest. 

The parameter $f_2$ produces a qualitatively similar impact in the $M_z$=[-23,-22] bin but with the opposite trend: higher $f_2$ suppresses small-scale clustering. This is because raising $f_2$ shifts the $f_{\rm sat}$ curve toward fainter magnitudes, thereby lowering the satellite fraction for bright galaxies. In contrast, the faint bin ($M_z$=[-21,-20]) is largely insensitive to changes in $f_2$ ($\Delta f_2$ = 0.3–0.6), as the $f_{\rm sat}$ curve is nearly flat over this magnitude range.

Compared with $f_1$ and $f_2$, the influence of $f_3$ on $w_p$ is relatively weak in the magnitude bin $M_z$=[-23,-22], but becomes more pronounced in the bin $M_z$=[-21,-20]. This behavior can be interpreted as follows. The parameter $f_3$ sets the width of the $f_{\rm sat}$ curve (i.e., the dashed curves in Figure \ref{fsat}). Increasing $f_3$ makes the $f_{\rm sat}$ curve narrower, which sharpens the transition in $f_{\rm sat}$ from bright to faint magnitudes. However, varying $f_3$ barely changes the mean value of $f_{\rm sat}$ around the inflection point, which is determined by $f_2$. In this mock catalog, the true value of $f_2$ is -22.6, which falls within the bin $M_z$=[-23,-22]. Consequently, modifications to $f_3$ have only a small effect on the average satellite fraction, and thus on the clustering signal, in this magnitude range. 

In contrast, within the $M_z$=[-21,-20] bin, variations in $f_3$ lead to a more pronounced change in $f_{\rm sat}$. From the fiducial $f_3$=0.56, increasing it by $\Delta f_3$=0.4 (red curves in the bottom-right panel) results in $f_3$=0.96 and raises the mean satellite fraction to $f_{\rm sat}\sim0.313$, only slightly higher than the fiducial $f_{\rm sat}\sim0.3$. Consequently, the corresponding $w_p$ for $\Delta f_3$=0.4 is only modestly enhanced relative to the fiducial $w_{\rm p0}$. Conversely, reducing $f_3$ by $\Delta f_3$=-0.4 (blue curves in the bottom-right panel) gives $f_3$=0.16, which lowers $f_{\rm sat}$ to $\sim0.2$ for $M_z$=[-21,-20], significantly below the fiducial $f_{\rm sat}\sim0.3$. This explains both the diminished amplitude and the larger departure of $w_p$ for $\Delta f_3$=-0.4 from $w_{\rm p0}$, relative to the $\Delta f_3$=0.4 case. Overall, the $f_{\rm sat}$ parameters serve as an effective lever for tuning the clustering strength, especially on small scales.

\begin{figure*}[!ht]
\centering
\includegraphics[width=0.85\textwidth]{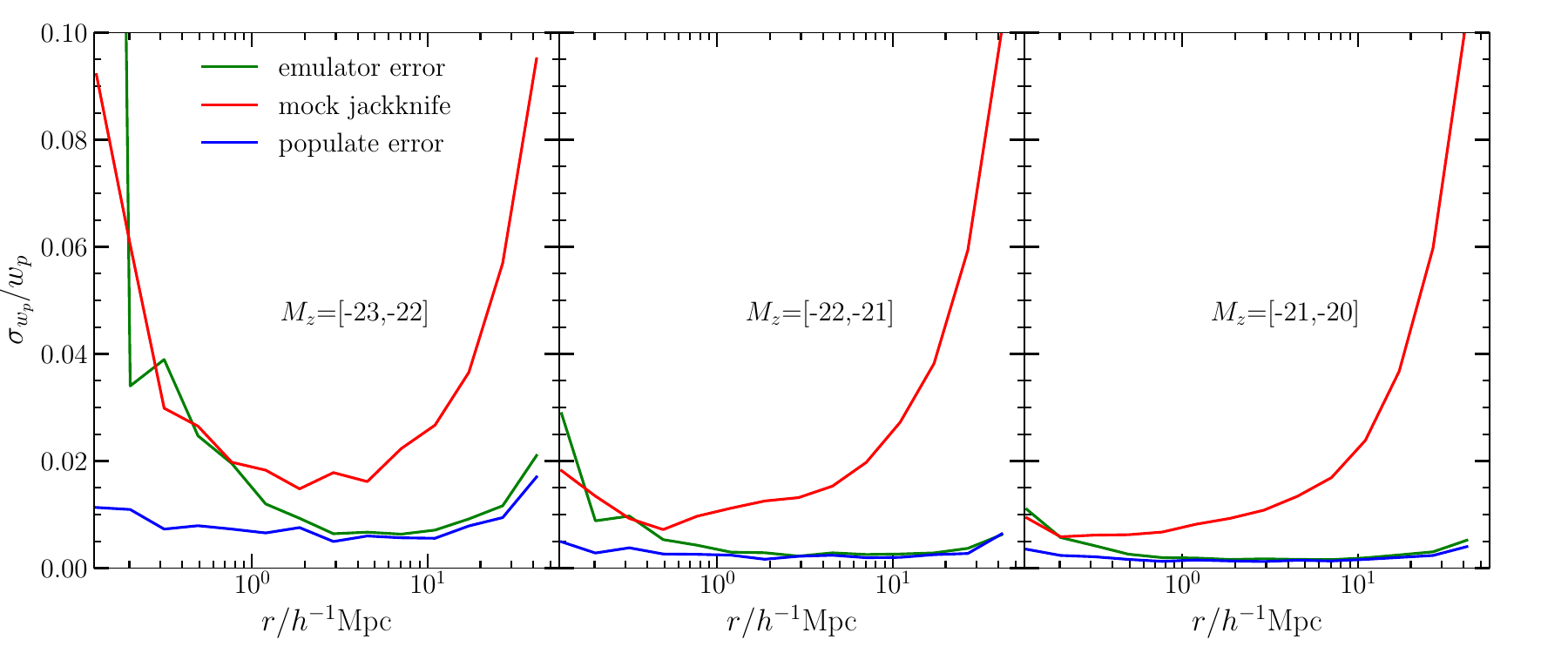}
\caption{Error in $w_p$ measurements normalized by $w_p$ of the Jiutian $M_{\rm peak}$ mock catalog. From left to right, the panels show $M_z=[-23,-22]$, $M_z=[-22,-21]$, and $M_z=[-21,-20]$. In each panel, jackknife error is indicated by red solid curve, and emulator error is indicated by green solid curve. The Possion error, computed by populating the subhalos with the same set of parameters multiple times, is also displayed in blue for comparison. }
\label{error-comp}
\end{figure*}

\subsection{Sobol sequence and emulator}
\label{sec:emulator}

Modeling $w_p$ with CS-SHAM requires populating galaxies using a given set of parameters and then evaluating galaxy clustering across various $M_z$ or stellar-mass bins. This procedure, however, becomes computationally demanding when applied within MCMC frameworks over large cosmological volumes. To mitigate this cost, we develop Gaussian process emulators that can rapidly predict $w_p$, thereby speeding up the MCMC analysis in this work. The emulator calibrated on the mock catalogs can also be employed to fit the observed DESI clustering in future studies. 

We first apply the Sobol sequence method \cite{Sobol1967} to draw 512 parameter combinations from the prior space specified in Table \ref{tab:prior}. The Sobol sequence is widely used to sample parameter spaces in cosmological simulation suites \cite{Villaescusa2020,DeRose2023,Chen2025a,Chen2025b}. It produces quasi-uniform samples in high-dimensional spaces via binary radical inversion, making it particularly well suited for MCMC applications.

\begin{table}[H]
\caption{prior parameter for Sobol sequence and MCMC}
\label{tab:prior}
\centering
\begin{tabular}{lccc}
\toprule
parameter &  mock  & SAM  & TNG-300 \\ \hline
$\sigma$ & [0.01,1.0]  & [0.01,0.5]  & [0.01,0.5]   \\
$f_1$& [0.05,0.21]  & [0.05,0.25] & [0.05,0.3] \\
$f_2$& [-24,-19]  & [10,13]  & [10,13] \\
$f_3$& [0.4,0.8]  & [0.7,1.2]  & [0.7,1.2]\\
\bottomrule
\end{tabular}
\end{table}

The parameters drawn from the Sobol sequence are subsequently employed to populate galaxies within subhalos, following the procedure described in Section \ref{sec:sham}. Using these Sobol-sequence parameters together with the corresponding $w_p$, we build Gaussian process emulators for each $r_p$ bin over all magnitude or stellar mass bins, following \cite{Zhai2019}. Gaussian process regression is a commonly used, non-parametric machine learning method that models the output variable (e.g., $w_p$ at fixed $r_p$) as a multivariate Gaussian distribution over different input variables (e.g., the CS-SHAM parameters), characterized by a chosen covariance (kernel) function. This kernel function encodes how the outputs are correlated with one another. As in \cite{Zhai2019}, we adopt a combination of a squared exponential kernel and a Matérn kernel. The posterior distribution of the emulator predictions is then obtained from the kernel specification and the training data. Once the Gaussian process emulators are constructed, we perform MCMC analyses using the {\tt emcee} package \cite{Foreman-Mackey2013}.

The $\chi^2$ statistic we use in the MCMC analysis is given by
\begin{equation} \label{eq:likelyhood}
\chi^2=\Sigma_i\,(w_{p,i}^{\rm obs}-w_{p,i}^{\rm model})^{\rm T}C_{i}^{-1}(w_{p,i}^{\rm obs}-w_{p,i}^{\rm model}),
\end{equation}
where $i$ indexes the $i$th $M_z$ or stellar-mass bin, and $C$ is the total covariance matrix, $C=C_{\rm jack}+C_{\rm emu}$. Here, $C_{\rm jack}$ is the jackknife covariance matrix of the observed $w_p$, while $C_{\rm emu}$ represents the covariance arising from the emulator uncertainty. We define the emulator error as the standard deviation of the relative difference between the emulator-predicted $w_p$ and the $w_p$ measured directly by populating galaxies using a given set of parameters $\theta$:
\begin{equation} \label{eq:likelyhood}
\sigma\!\left(\frac{w_p^{\rm emu}(\theta)-w_p^{\rm pop}(\theta)}{w_p^{\rm pop}(\theta)}\right).
\end{equation}
Following \cite{Zhai2019}, we incorporate this emulator error only along the diagonal of $C_{\rm emu}$ and set all off-diagonal terms to zero.

In Figure \ref{error-comp}, we show the diagonals of  
$C_{\rm jack}$ and $C_{\rm emu}$ (i.e., the jackknife error and emulator error), normalized by $w_p$ for three $M_z$ bins from the Jiutian $M_{\rm peak}$ mock catalog. The jackknife error is computed by separating the galaxy samples into 64 subsamples  according to position using K-means clustering and measuring $w_p$ with one subsample omitted at a time. For comparison, we also include the Poisson error, computed by populating the subhalos multiple times with the same set of parameters. In the bin of $M_z=[-23,-22]$, the jakknife error is large ($\sim$ 10\%) at both small and large scales, with a smaller value ($\sim$ 2\%) at intermediate scales. The emulator error is comparable to jakknife error for  scales $r<1\hmpc$ and similar to Poisson error (1\%$\sim$ 2\%) on $r>2\hmpc$ (much smaller than the jackknife error). This trend of emulator error is also observed in the other two $M_z$ bins, with the emulator error on large scales remaining below 0.5\%. We note that our emulator error is smaller compared to those reported for cosmological emulators \cite{Zhai2019}, which could be attributed to the fixed cosmology used in this work. 


\begin{figure*}[!ht]
\centering
\includegraphics[width=0.85\textwidth]{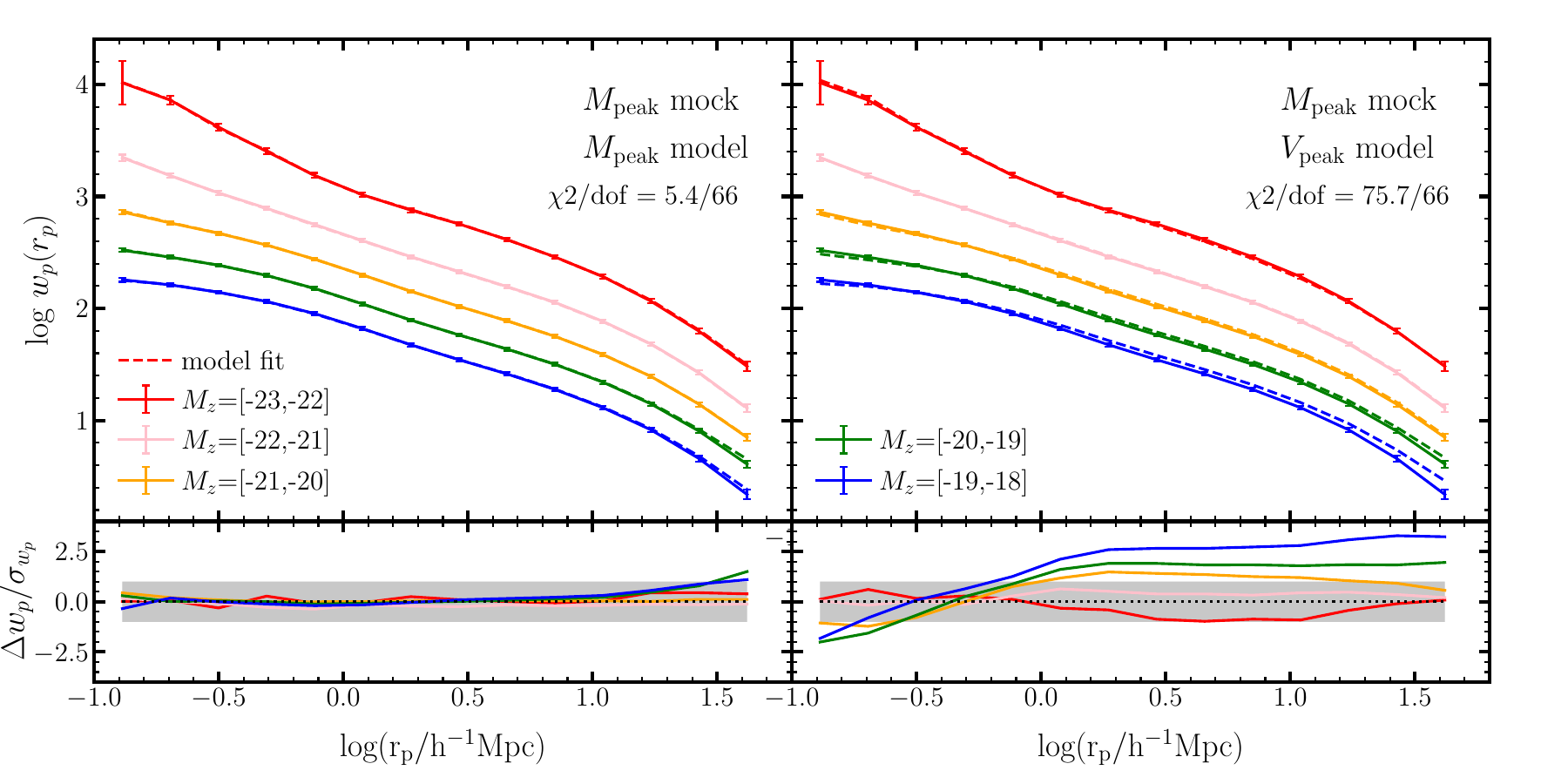}
\includegraphics[width=0.85\textwidth]{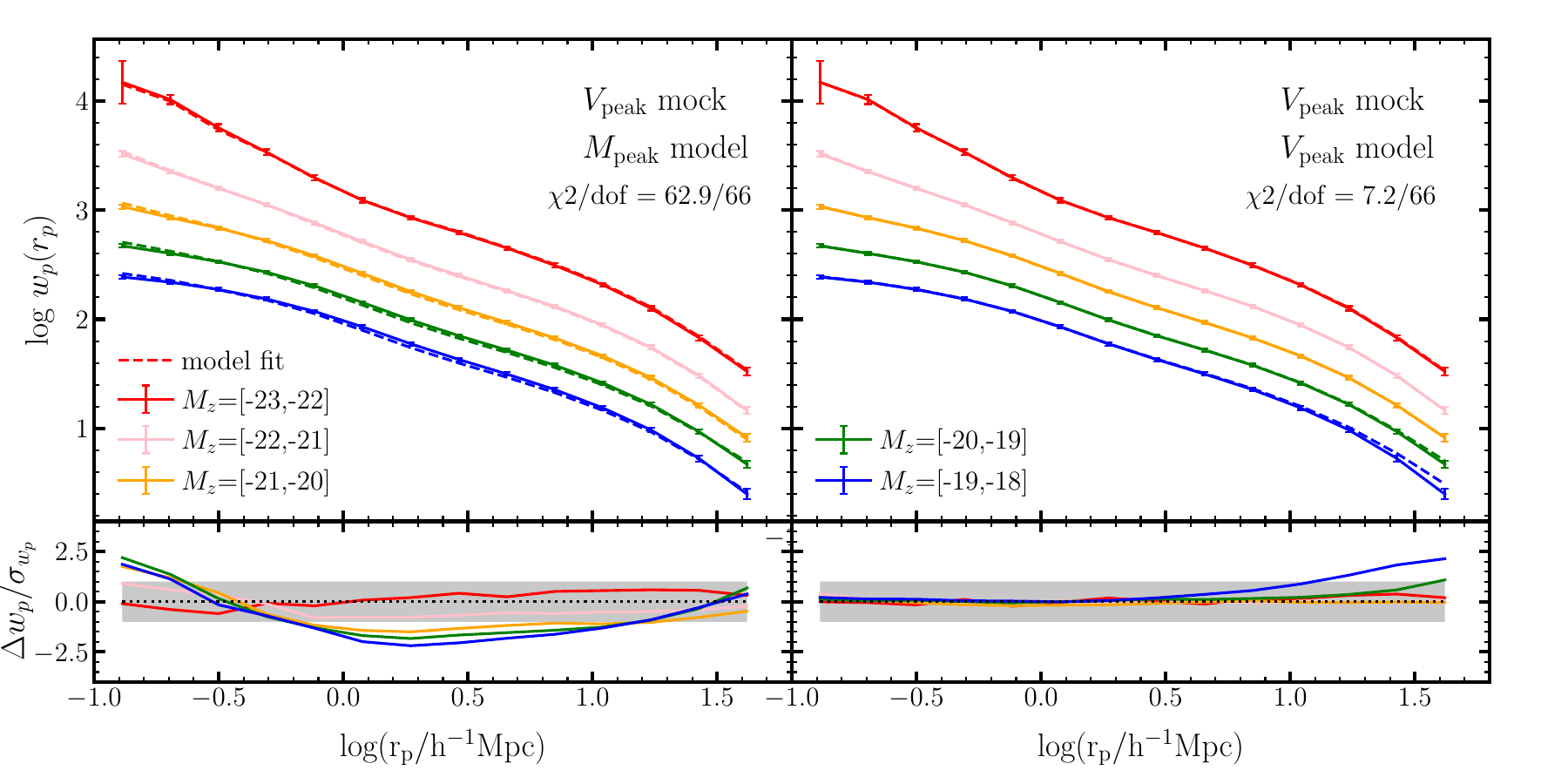}
\caption{Best-fit $w_p$ results obtained by applying the $M_{\rm peak}$ model (left) and the $V_{\rm peak}$ model (right) to the $M_{\rm peak}$ mock (top) and the $V_{\rm peak}$ mock (bottom). The colored dashed curves represent the best-fit $w_p$ for five $M_z$ bins spanning $M_z=-23$ to $M_z=-18$, while the solid curves with error bars show the corresponding direct measurements from the mocks. The error bars correspond to the square root of the diagonal elements of the covariance matrix, which is computed as the quadratic sum of the jackknife uncertainty and the emulator uncertainty, with an additional 3\% error included. In each panel, for visual clarity, each absolute magnitude bin is offset by 0.2 dex, with no shift applied to the faintest bin. The lower panels present the residuals between the best-fit and the true $w_p$, divided by the square root of the diagonal elements of the covariance matrix. The shaded band denotes the $1\sigma$ region. The $\chi^2$ value for each model is displayed in the top right corner.}
\label{result_mockclustering}
\end{figure*}

\begin{figure*}[!ht]
\centering
\includegraphics[width=0.45\textwidth]{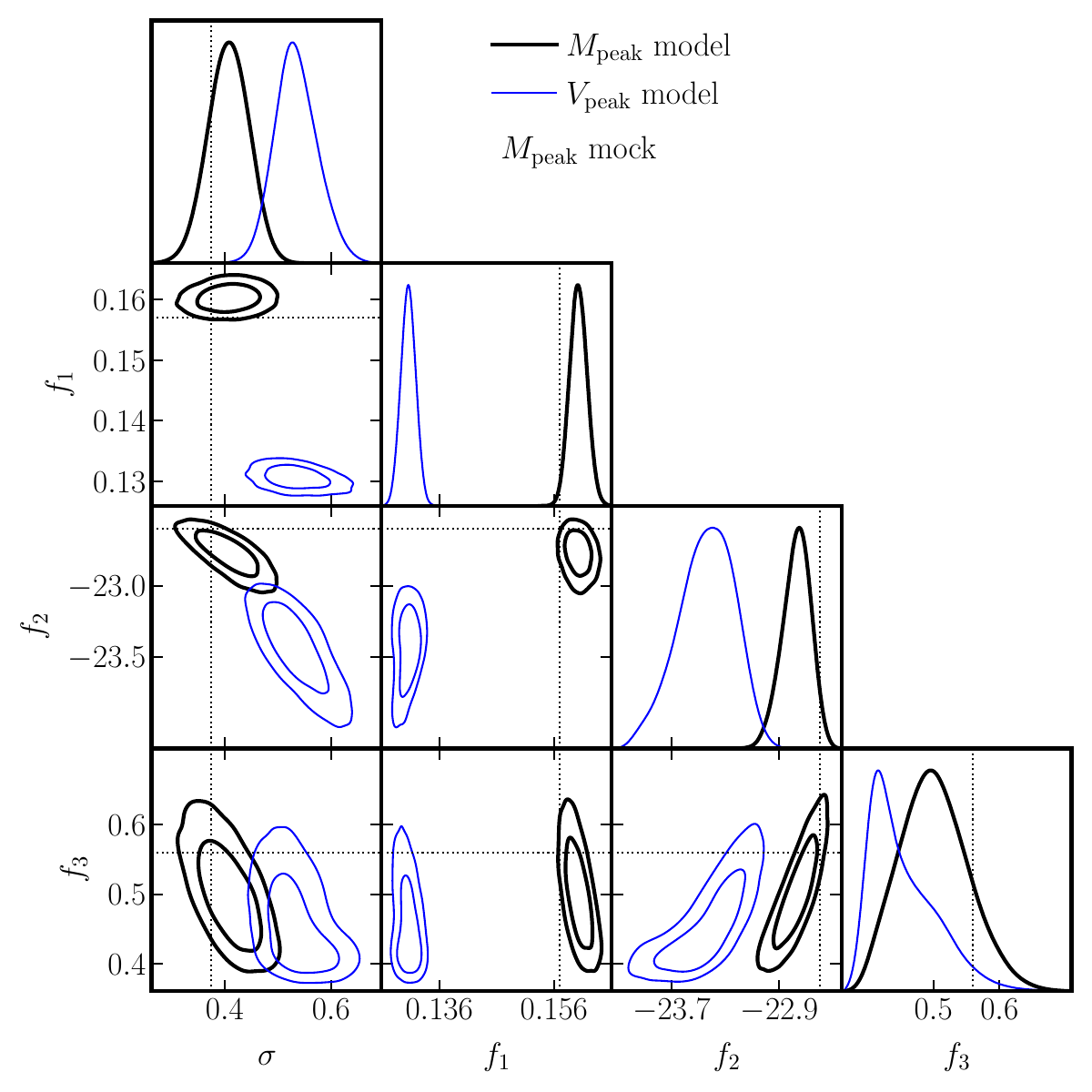}
\includegraphics[width=0.45\textwidth]{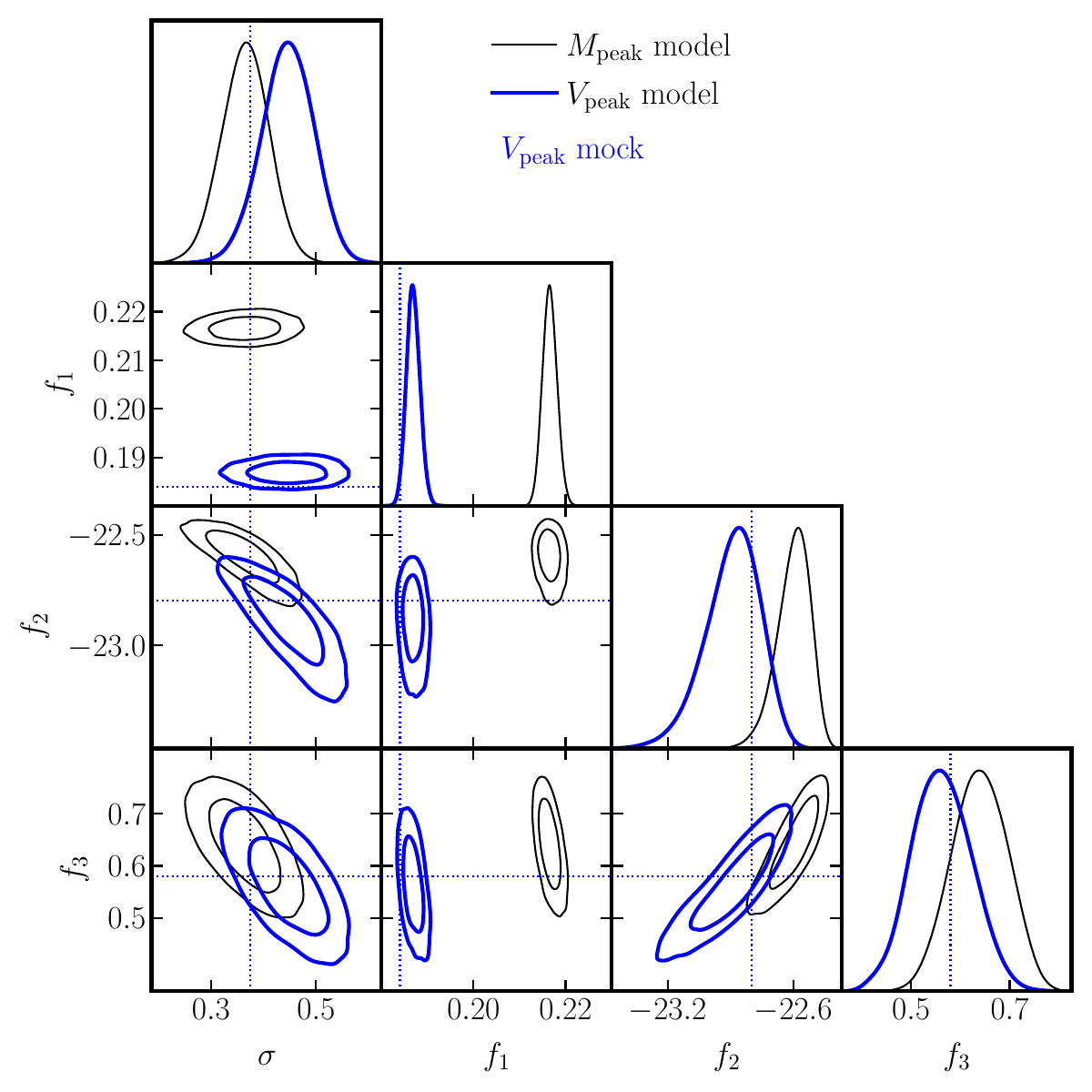}
\caption{Parameter constraints for the $M_{\rm peak}$ (left panel) and $V_{\rm peak}$ (right panel) mock catalogs, obtained using the $M_{\rm peak}$ model (black) and the $V_{\rm peak}$ model (blue), respectively. The true parameter values are shown as dotted horizontal and vertical lines. }
\label{result_mockparam}
\end{figure*}

\begin{figure*}[!ht]
\centering
\includegraphics[width=0.85\textwidth]{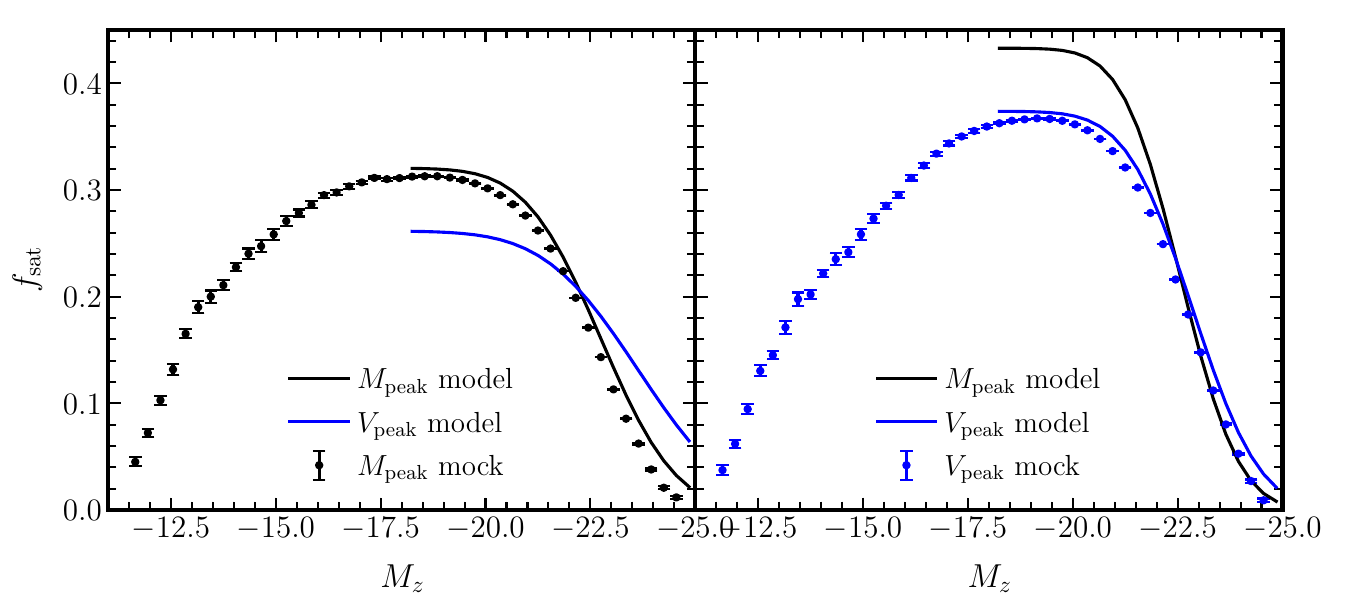}
\caption{$f_{\rm sat}$ as a function of $M_z$ using the best-fit parameters from each model applied to each mock. The results for the $M_{\rm peak}$ mock and the $V_{\rm peak}$ mock are displayed in the left and right panels, respectively. In each panel, the best-fit $f_{\rm sat}$ from the $M_{\rm peak}$ model (black solid line) and from the $V_{\rm peak}$ model (blue solid line) are shown for $M_z < -18$. For reference, the true $f_{\rm sat}$ measured directly from the mock is plotted as colored points with error bars.}
\label{result_mockfsat}
\end{figure*}

\section{Results}
\label{sec:results}

In this section, we present the results of galaxy clustering fitting and parameter constraints for the Jiutian mocks, ELUCID SAM, and TNG-300. For all the galaxy samples, we adopt the 4-parameter CS-SHAM model: one parameter for the scatter in the stellar-to-subhalo mass relation, and three parameters for $f_{\rm sat}$.

\begin{figure*}[!ht]
\centering
\includegraphics[width=0.85\textwidth]{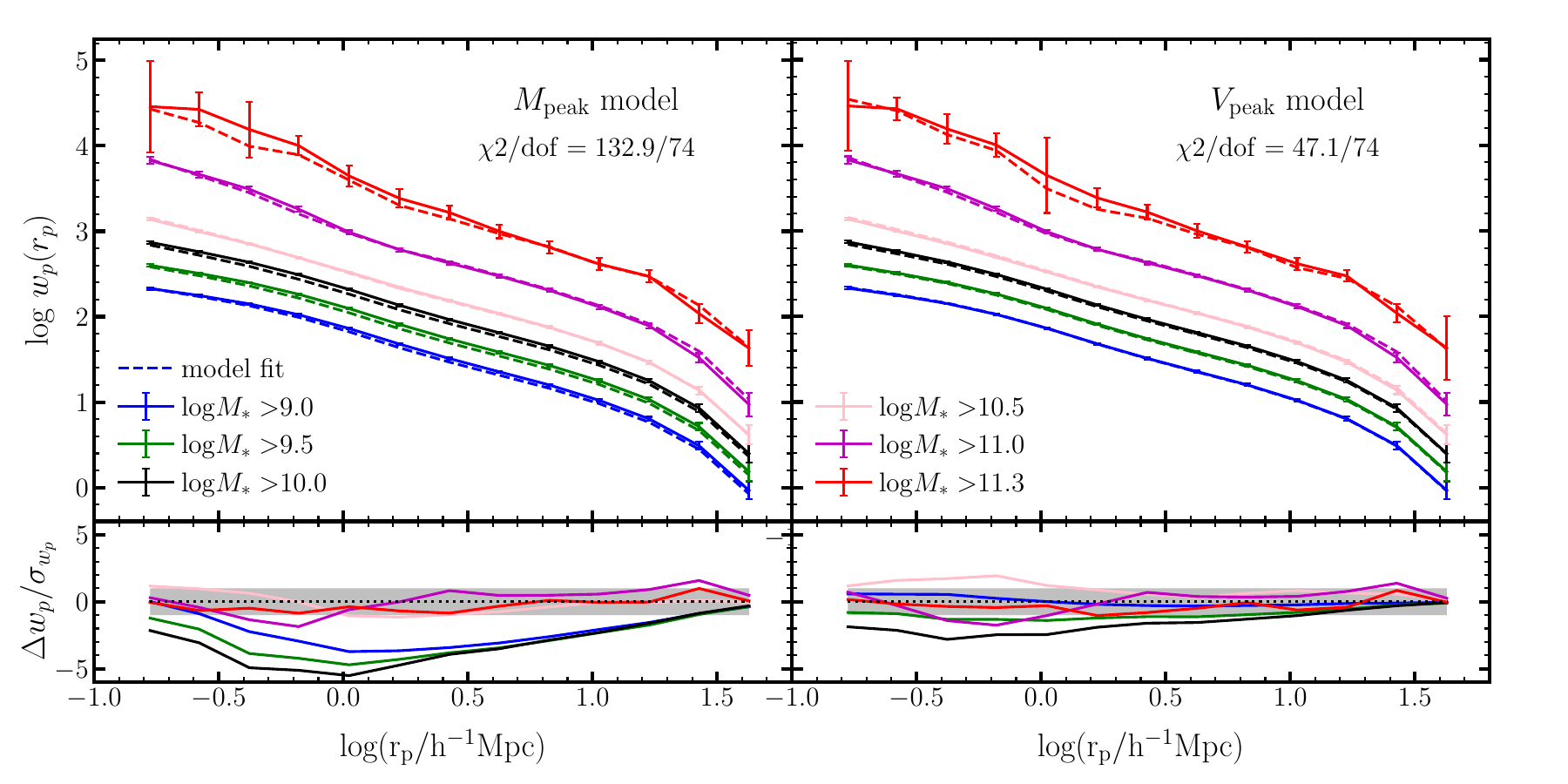}
\includegraphics[width=0.85\textwidth]{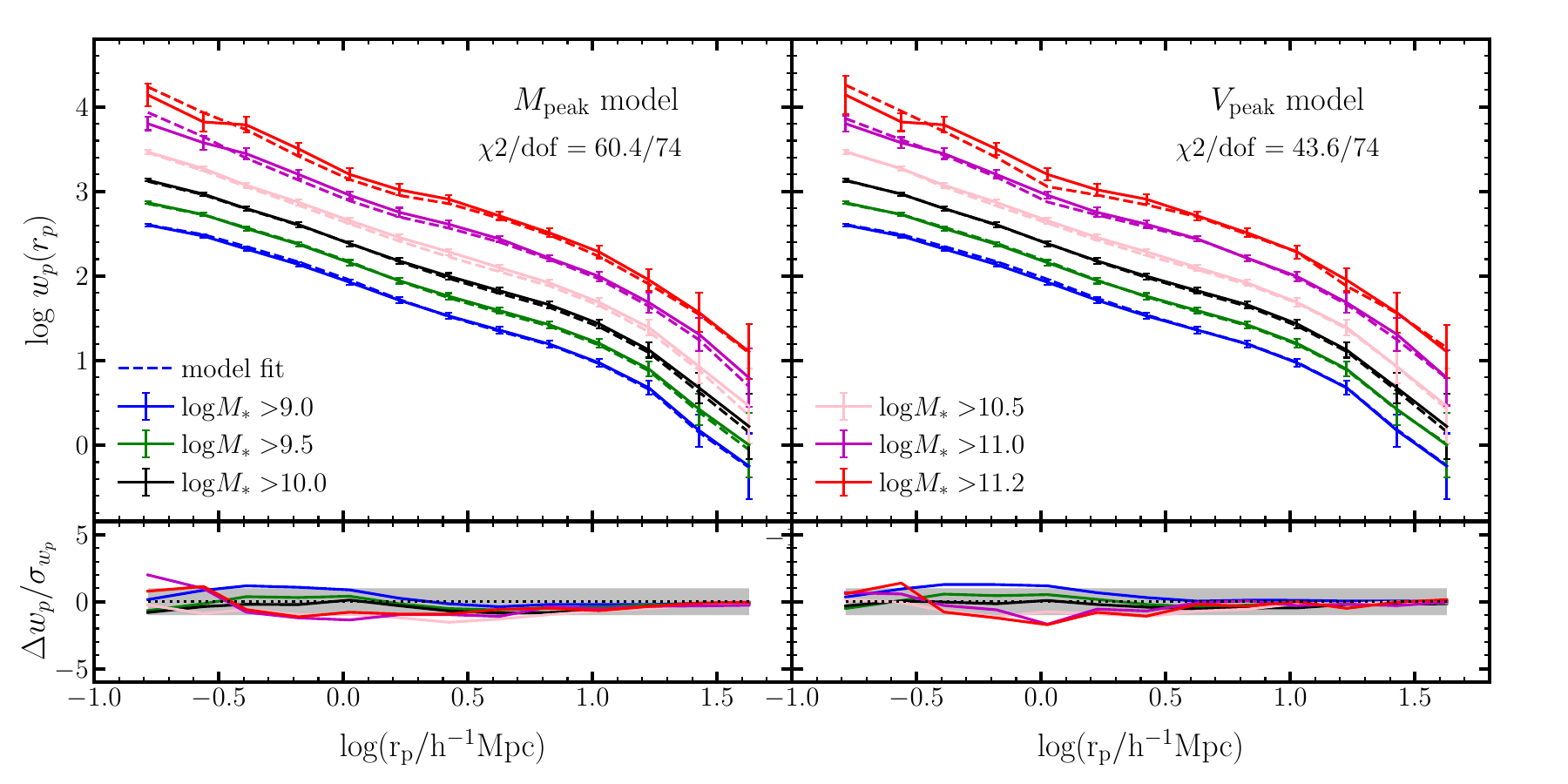}
\caption{Best-fit $w_p$ results derived from fitting the $M_{\rm peak}$ model (left) and the $V_{\rm peak}$ model (right) to the ELUCID SAM (top) and TNG-300 (bottom) mock catalogs, respectively.  
In each panel, the solid colored curves with error bars show the measured $w_p$ for six stellar-mass–threshold samples, while the colored dashed curves indicate the corresponding best-fit predictions. The error bars are given by the square root of the diagonal elements of the covariance matrix, which includes, in quadrature, both jackknife and emulator errors. For visual clarity, each stellar mass threshold is offset by 0.2 dex, except for the lowest-mass bin, whose $w_p$ is plotted without any shift. The $\chi^2$ value for each fit is displayed in the upper-right corner of each panel. The smaller subpanels at the bottom report the residuals between the best-fit model and the true $w_p$, normalized by the square root of the covariance-matrix diagonal. The shaded band denotes the $1\sigma$ range.}
\label{sam_wp}
\end{figure*}

\subsection{Fitting Jiutian mock clustering}
\label{sec:fitmock}

The Jiutian mocks in \cite{Gu2024} consist of three catalogs, each constructed using basic SHAM with a fixed $\sigma_{\rm log L}$ based on subhalo mass indicators $M_{\rm peak}$, $V_{\rm max}$, and $V_{\rm peak}$. We make use of two of them and refer to them as the $M_{\rm peak}$ mock and $V_{\rm peak}$ mock, respectively. We apply CS-SHAM using these two mass indicators, which we refer to as the $M_{\rm peak}$ model and $V_{\rm peak}$ model in the following. For each mock, we fit both the models and compare their performances, providing insights into the effects of using different mass indicators in CS-SHAM modeling. 

We use the subhalos in Jiutian lightcone catalog within the redshift range $z=[0,0.3]$ and measure subhalo mass function. This approach differs slightly from the subhalo mass functions used in generating the mock galaxy catalogs, which are directly measured from the simulation box at specific snapshots. To account for this and the difference in subhalo lightcone used, we add an additional 3\% error to the diagonal terms of the covariance matrix. We also tested additional errors of 1\% and 5\% and found that these only affect the values of $\chi^2$ and the width of parameters posteriors, with the best-fit $w_p$ and parameters remains unchanged. In addition, galaxy luminosity functions, in term of $z$-band absolute magnitude $M_z$, are measured in the same redshift bins. We then fit $w_p$ for five $M_z$ bins in the mocks, ranging  from $M_z=-23$ to $M_z=-18$. 

Figure \ref{result_mockclustering} presents the best-fit $w_p$ for each model applied to the $M_{\rm peak}$ mock (top) and the $V_{\rm peak}$ mock (bottom), alongside the original clustering measured in these mocks. For the $M_{\rm peak}$ mock, the $M_{\rm peak}$ model accurately reproduces the mock clustering across all five bins. In contrast, the best-fit $w_p$ from the $V_{\rm peak}$ model exhibits small offset within $1\sigma_{w_p}$ from the mock clustering for the two brightest bins, with larger discrepancies for the three fainter bins. Nevertheless, it is still considered an acceptable prediction with $\chi^2/{\rm dof}=1.15$. For the $V_{\rm peak}$ mock, the $V_{\rm peak}$ model yields the most consistent $w_p$, while the $M_{\rm peak}$ model shows larger discrepancies within $2\sigma_{w_p}$ and $\chi^2/{\rm dof}=0.95$. Overall, CS-SHAM using any of the models is able to reasonably recover $w_p$ in the mocks.

Figure \ref{result_mockparam} presents the parameter constraints for both models when applied to the $M_{\rm peak}$ mock (left) and the $V_{\rm peak}$ mock (right). For the $M_{\rm peak}$ mock, the $M_{\rm peak}$ model is able to reasonably recover the true input parameters that the true ones lie within $2\sigma$ of the inferred posterior distributions. In contrast, the parameters inferred with the $V_{\rm peak}$ model deviate substantially from the true values, preferring higher $\sigma$ and lower values of $f_1$, $f_2$ and $f_3$. For the $V_{\rm peak}$ mock, the $V_{\rm peak}$ model provides a more accurate recovery of the true parameters within $2\sigma$. To reproduce the $w_p$ measured from this mock, the $M_{\rm peak}$ model requires a much larger $f_1$ (close to the upper bound of our prior), implying a higher satellite fraction.

Since the parameters can be degenerate, it is more informative to examine the $f_{\rm sat}$ curve directly. Figure \ref{result_mockfsat} presents the best-fit $f_{\rm sat}$ curves and compares them with the true curves in the mocks. This comparison offers an additional assessment of how well the model recovers the true satellite fraction, beyond what can be inferred from the posterior distributions of the model parameters, especially because the true $f_{\rm sat}$ may differ from the error-function form assumed in Equation \ref{eq:fsat2}. For both mock catalogs, the model that uses the same subhalo mass indicator as the one employed to construct the mock yields a more accurate recovery of the true $f_{\rm sat}$. However, if an alternative mass indicator is adopted, the inferred $f_{\rm sat}$ can differ substantially from the true values, even though the galaxy clustering is still accurately reproduced.

\begin{figure*}[!ht]
\centering
\includegraphics[width=0.45\textwidth]{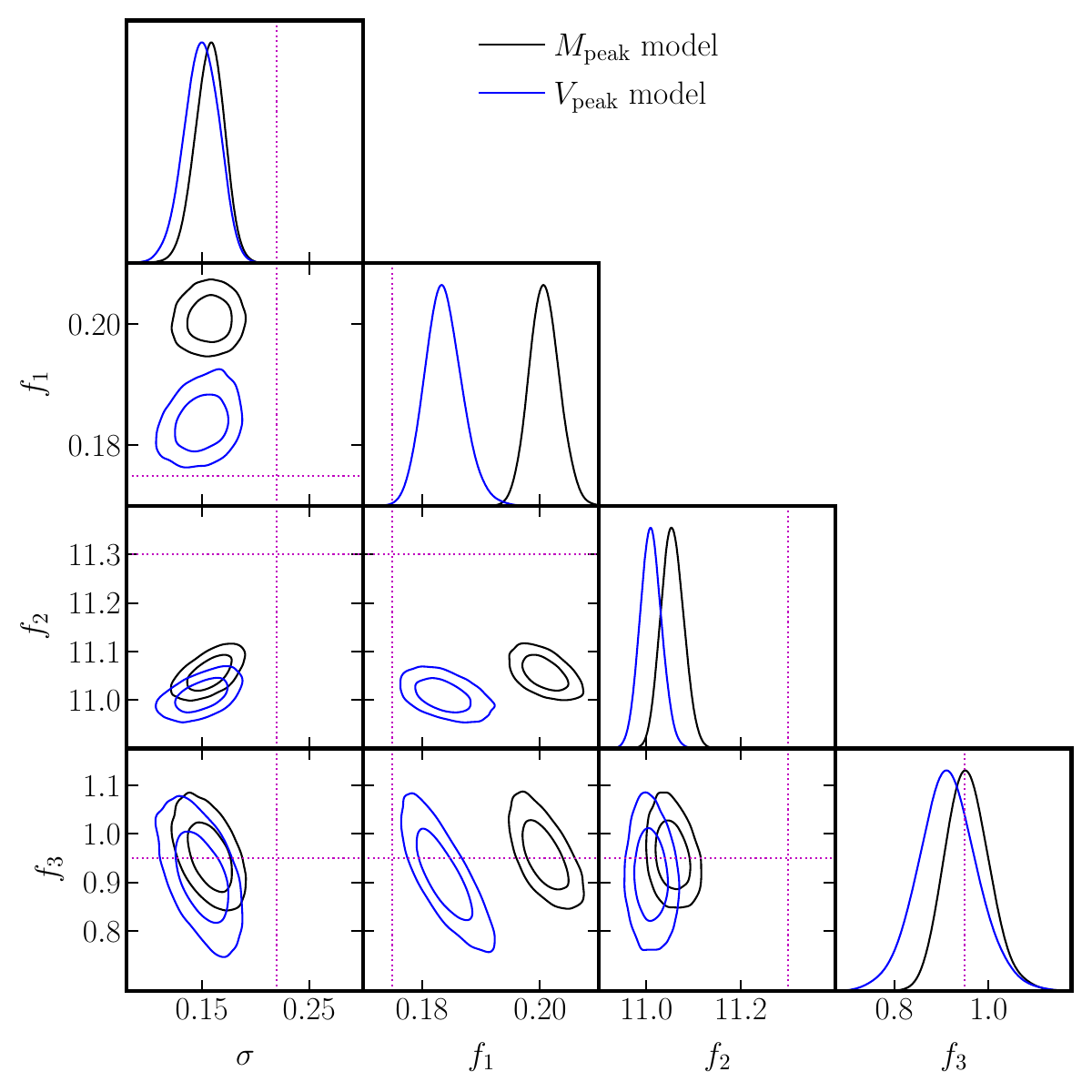}
\includegraphics[width=0.45\textwidth]{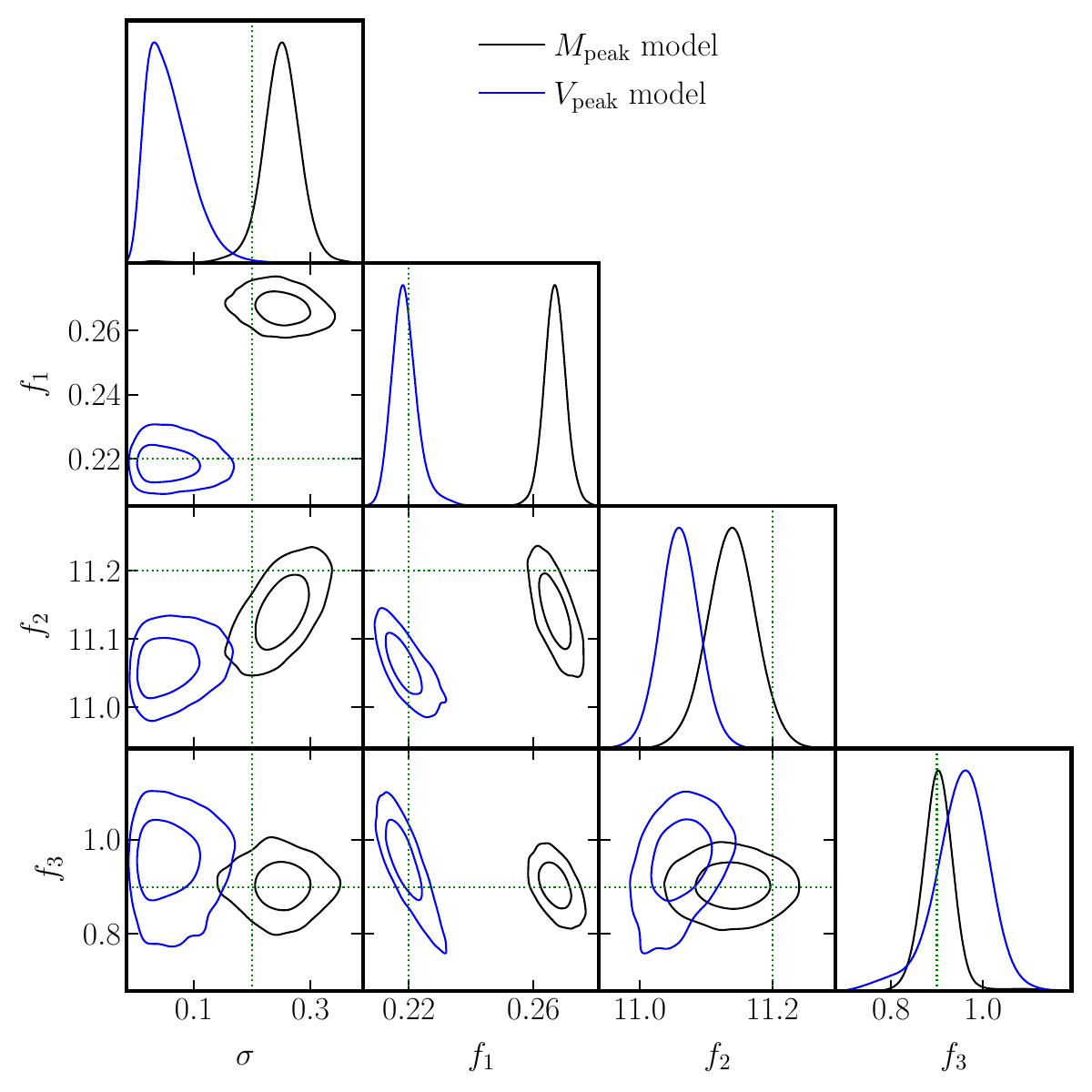}
\caption{Parameter constraints for the ELUCID SAM (left panel) and TNG-300 (right panel) using the $M_{\rm peak}$ and $V_{\rm peak}$ models. The true parameter values are shown as dotted horizontal and vertical lines. }
\label{sam_dist}
\end{figure*}

\begin{figure*}[!ht]
\centering
\includegraphics[width=0.40\textwidth]{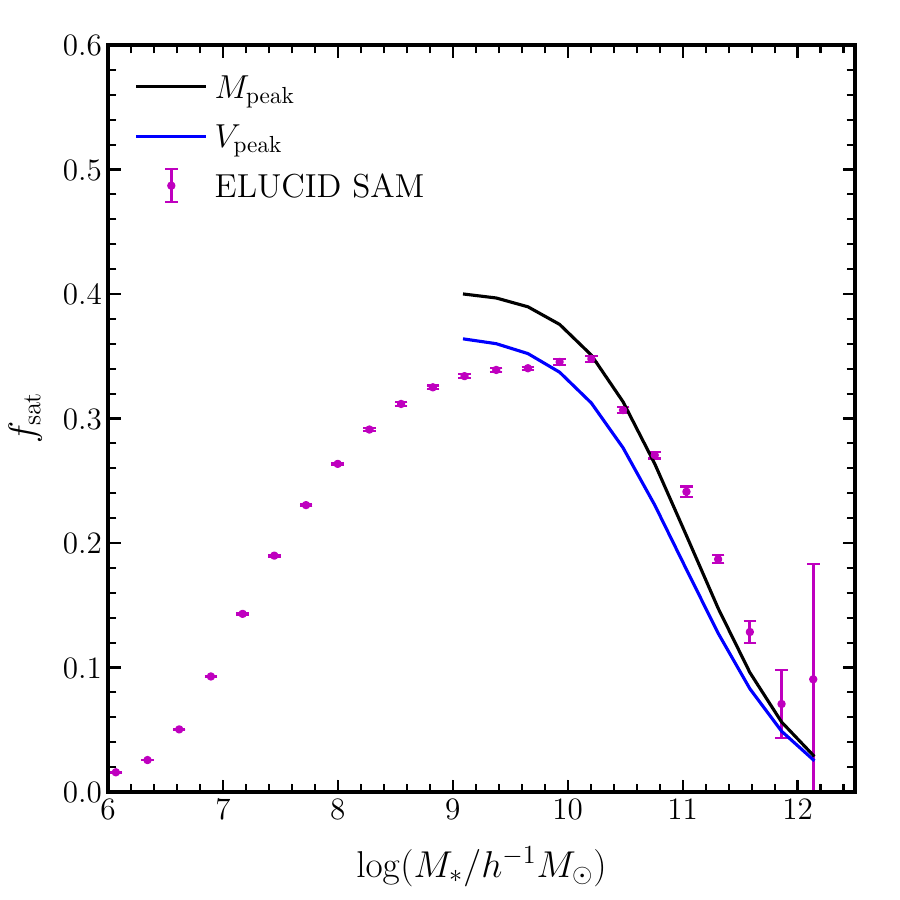}
\includegraphics[width=0.40\textwidth]{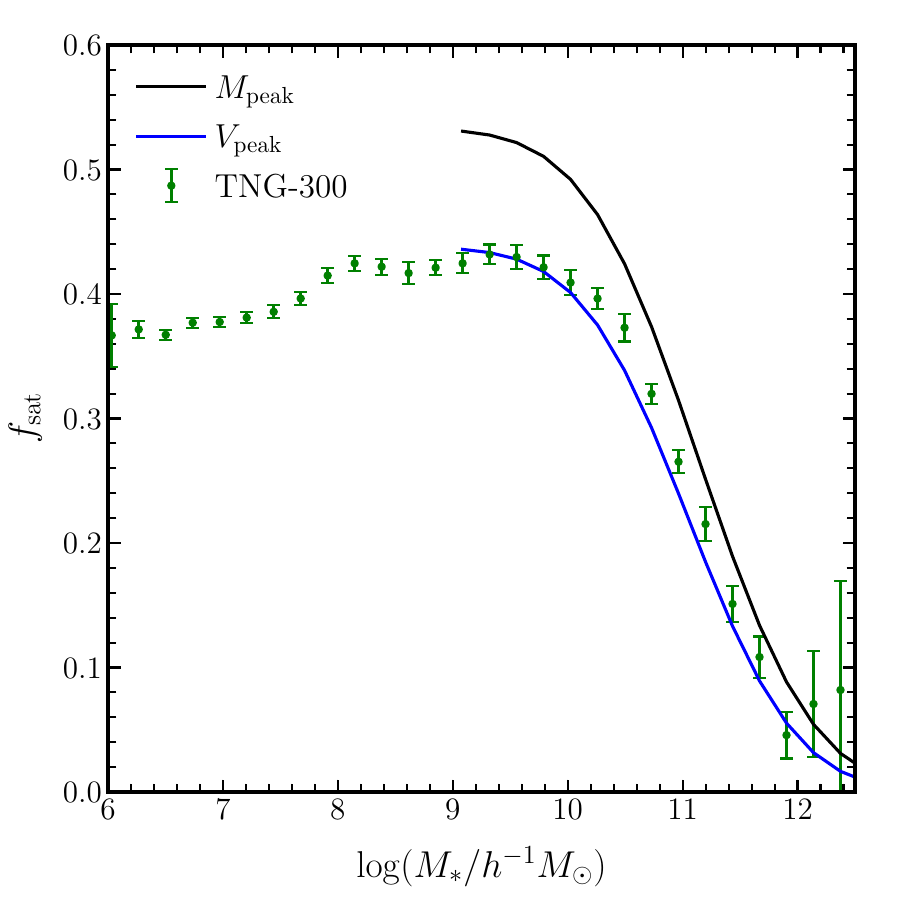}
\caption{Best-fit $f_{\rm sat}$ as a function of stellar mass for the $M_{\rm peak}$ and $V_{\rm peak}$ models. The left panel shows results for the ELUCID SAM, and the right panel for TNG-300. In each panel, the dots  with error bars indicate the true $f_{\rm sat}$.  }
\label{tng_fsat}
\end{figure*}

\subsection{Fitting ELUCID SAM and TNG-300 clustering}
\label{sec:fitsam}

In this section, we apply our model to the ELUCID SAM and TNG-300 samples introduced in Section \ref{sec:SAMhydro}. We partition each of these two samples into six stellar-mass threshold bins and measure $w_p$ for every bin. The CS-SHAM model is then fit jointly across all bins. Figure \ref{sam_wp} shows the resulting fits for the $M_{\rm peak}$ and $V_{\rm peak}$ models (with the top panels corresponding to ELUCID SAM and the bottom panels to TNG-300). The stellar-mass threshold bins are indicated in the legends. Solid curves with error bars depict the original clustering measurements, while the dashed curves show the corresponding best-fit CS-SHAM predictions. The small subpanels beneath each main panel display the residuals, normalized by the square root of the diagonal elements of the covariance matrix. For both ELUCID SAM and TNG-300, the $V_{\rm peak}$ model yields slightly better overall performance with a lower $\chi^2$ (displayed in the top-right corner of each panel). The best-fit $w_p$ predictions closely follow the true measurements and lie mostly within the 1$\sigma$ uncertainties across all bins. The $M_{\rm peak}$ model exhibits somewhat higher $\chi^2$ and slightly underestimates clustering in low stellar mass bins for ELUCID SAM and in high stellar mass bins for TNG-300, respectively.

\begin{figure*}[!ht]
\centering
\includegraphics[width=0.85\textwidth]{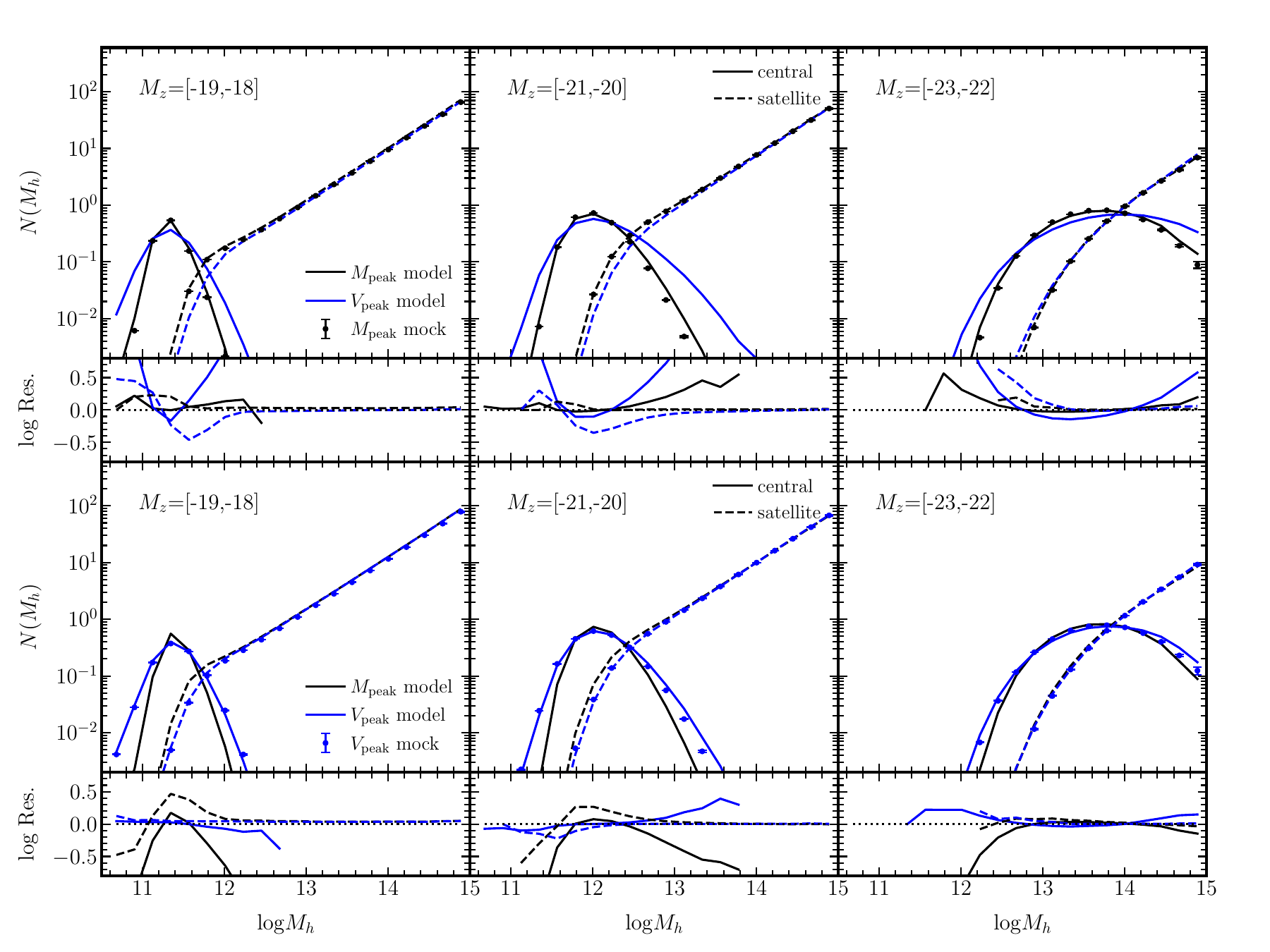}
\caption{Halo occupation for central and satellite galaxies in three $M_z$ bins. The top panels show the measured HOD from the $M_{\rm peak}$ mock (black dotted) together with the best-fit model predictions based on $M_{\rm peak}$ and $V_{\rm peak}$ (black and blue curves, respectively). Central/satellite predictions are presented by solid/dashed curves. The third row shows the measured HOD from the $V_{\rm peak}$ mock (blue dotted) along with the corresponding best-fit predictions from the two models. The second and bottom rows show the residuals of the log occupation defined by $\log \rm Res. = {\rm log}N_{\rm model}- {\rm log}N_{\rm mock}$.}
\label{result_mockhod}
\end{figure*}

The left panel of Figure \ref{sam_dist} shows the parameter constraints derived from the two models for the ELUCID SAM. In both cases, the inferred value of $\sigma$, which describes the scatter in ${\rm log}M_*$ at fixed subhalo mass indicators, is underestimated. A plausible explanation is that $\sigma$ may actually vary with subhalo mass indicators or stellar mass, rather than remaining constant as assumed in our model. Introducing a mass-dependent $\sigma$ could alleviate this tension, but would require additional free parameters, which we defer to future work.
Both models overestimate $f_1$, although the $V_{\rm peak}$ model yields values closer to those measured from the SAM, whereas the $M_{\rm peak}$ model returns substantially higher estimates.
Neither model recovers the true value of $f_2$, where the true value in the ELUCID SAM is about 11.3.  For $f_3$, both models yield results that are consistent with the ELUCID SAM.
The right panel of Figure \ref{sam_dist}  presents the constrained model parameters for TNG-300. The $M_{\rm peak}$ model yields $\sigma$ values closer to the true value in the TNG-300. The $V_{\rm peak}$ model successfully recovers the true value of $f_1$, whereas the $M_{\rm peak}$ model significantly overpredicts it. As in the SAM case, $f_2$ are underpredicted in all models due to the absence of higher stellar mass threshold bins. 

To further illuminate the global trends, we show the $f_{\rm sat}$ curves directly in Figure \ref{tng_fsat}. The left panel compares the predictions from the ELUCID SAM, using its best-fit parameters, with the true $f_{\rm sat}$ values. The $M_{\rm peak}$ model slightly overestimates $f_{\rm sat}$ at ${\rm log}M_*\sim 9$, but matches reasonably well for ${\rm log}M_*>10$. By contrast, the $V_{\rm peak}$ model shifts the $f_{\rm sat}$ distribution somewhat toward lower stellar masses relative to the SAM. The right panel shows the TNG-300 results: in this case, the $M_{\rm peak}$ model requires a significantly higher $f_{\rm sat}$ to reproduce the TNG-300 $w_p$, whereas the $V_{\rm peak}$ model yields a more realistic $f_{\rm sat}$ that closely tracks the true values. These trends suggest that although tuning $f_{\rm sat}$ can lead to an accurate description of galaxy clustering, the inferred $f_{\rm sat}$ does not necessarily coincide with the true values.

\section{What galaxy-halo connections?}
\label{sec:HOD-CLF}

Within our modified CS-SHAM framework, which contains only four free parameters, we have shown that galaxy clustering can be accurately reproduced across a broad range of luminosity and stellar mass bins, regardless of whether $M_{\rm peak}$ or $V_{\rm peak}$ of subhalos is adopted for abundance matching. However, the resulting $f_{\rm sat}$ does not necessarily match the true values, and therefore should be regarded only as a set of phenomenological tuning parameters. In this section, we revisit the traditional galaxy–halo connection frameworks, where galaxy clustering was widely used to constrain HOD and CLF/CSMF models, in order to assess which aspects can be tightly constrained.

\begin{figure*}[!ht]
\centering
\includegraphics[width=0.98\textwidth]{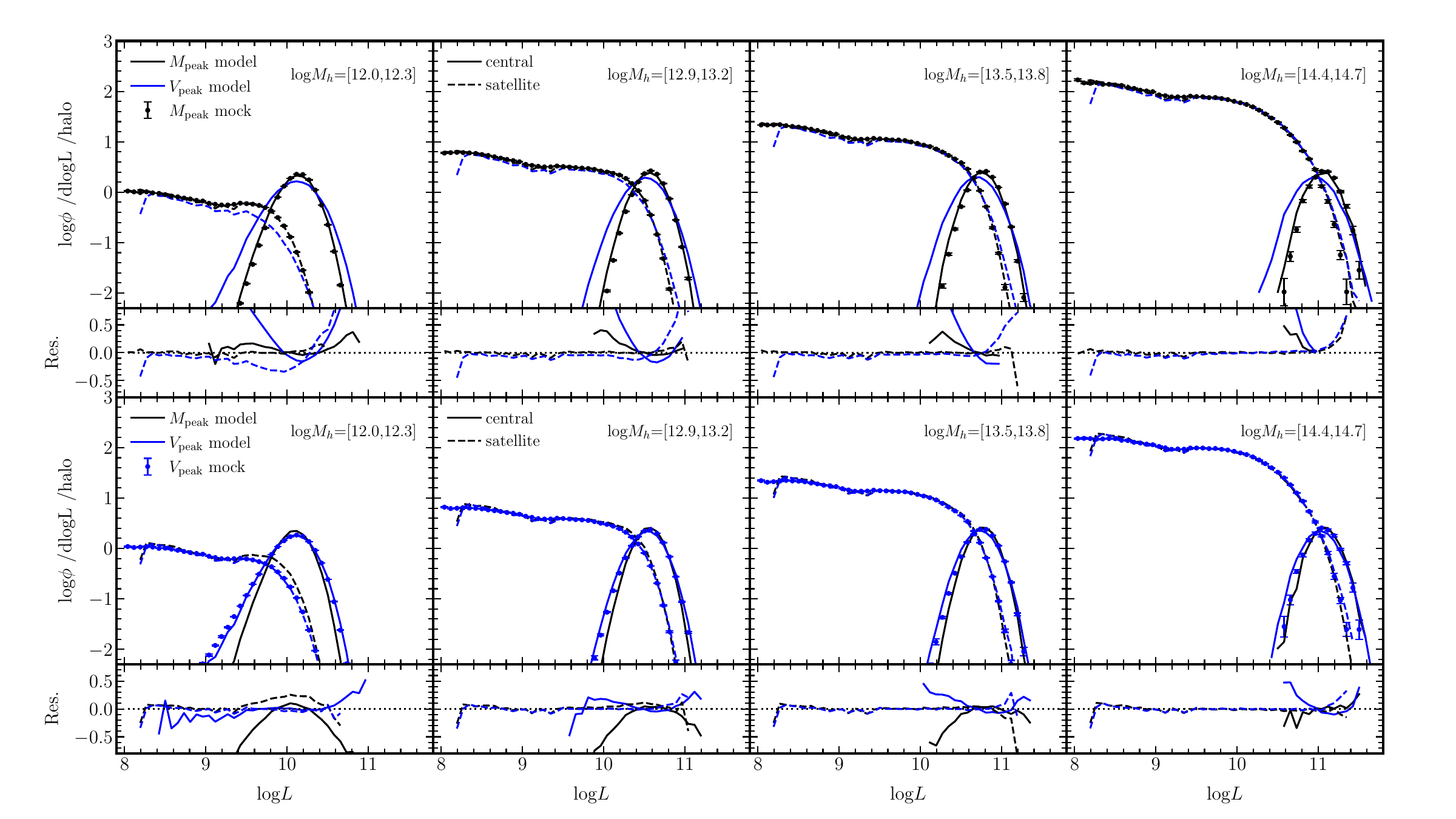}
\caption{Conditional luminosity function (CLF) of central and satellite galaxies in four bins of halo mass. Direct measurements are shown as dots with error bars, while the best-fit predictions from the $M_{\rm peak}$ and $V_{\rm peak}$ models are indicated by black and blue curves, respectively. Centrals/satellites are indicated by solid/dashed curves. The top panels present results for $M_{\rm peak}$ and the third row for $V_{\rm peak}$. The second and bottom rows show the residuals.}
\label{result_mockclf}
\end{figure*}

\subsection{HOD and CLF for Jiutian mocks}
\label{sec:HOD_Jiutian}

We begin by validating the HOD by comparing the predictions from our CS-SHAM model with the true HOD measured from the Jiutian mock catalogs, as shown in Figure \ref{result_mockhod}. Using the best-fit parameters, we can easily predict the HOD for any bin within the considered $M_z$ interval. For illustration, we present three representative $M_z$ bins corresponding to bright ($M_z$ = [-23, -22]), intermediate ($M_z$ = [-21, -20]), and faint ($M_z$ = [-19, -18]) galaxies. As expected, when CS-SHAM employs the same subhalo mass proxy that was used to construct the mock, the resulting HOD matches the true mock HOD very well. On the other hand, unlike the $f_{\rm sat}$ curves—which show a pronounced variation when the alternative mass proxy is adopted—the HOD still yields a good agreement between the best-fit and true values in this case, especially in massive halos.

For instance, in the top panels, the $M_{\rm peak}$ model closely matches both the central and satellite HOD of the $M_{\rm peak}$ mock, whereas a slight discrepancy is visible for the $V_{\rm peak}$ model. Considering first the central galaxies, the $V_{\rm peak}$ model yields a somewhat broader range of host halo masses at fixed $M_z$, though it still recovers the peak host halo mass without systematic offset, mainly because of the constant $\sigma_L$ we assume. Note that, by construction, $\sigma_L$ is fixed with respect to each model’s own subhalo mass indicator, but this no longer holds when we switch to a different mass indicator. Even so, modest changes in $\sigma_L$ have little impact on galaxy clustering and therefore do not hinder future cosmological applications.

Next, we turn to the satellite galaxies. Although the $V_{\rm peak}$ model yields satellite populations that very closely match the true values in massive halos with $M > 10^{13}\hmsun$, clear deviations emerge in halos of lower mass. In the faintest bin, the predicted satellite counts from the two models can differ from the true values by nearly $\log \rm Res \sim 0.5$, but this discrepancy decreases to about $\log \rm Res \sim 0.04$ for massive halos. Because low-mass halos exhibit an almost flat bias for masses below $10^{13.0}\hmsun$ \cite{Sheth2001}, whether low-mass galaxies are labeled as centrals or satellites within these halos does not substantially affect the clustering strength. This, in turn, leads to the larger deviation in $f_{\rm sat}$ for faint galaxies seen in Figure \ref{result_mockfsat}.

In the lower panels, we present results for the $V_{\rm peak}$ Jiutian mocks. The behavior is very similar to that in the upper panels; however, in this case the $M_{\rm peak}$ model exhibits a noticeable discrepancy. The underlying reasons for this discrepancy are analogous to those discussed above.

\begin{figure*}[!ht]
\centering
\includegraphics[width=0.85\textwidth]{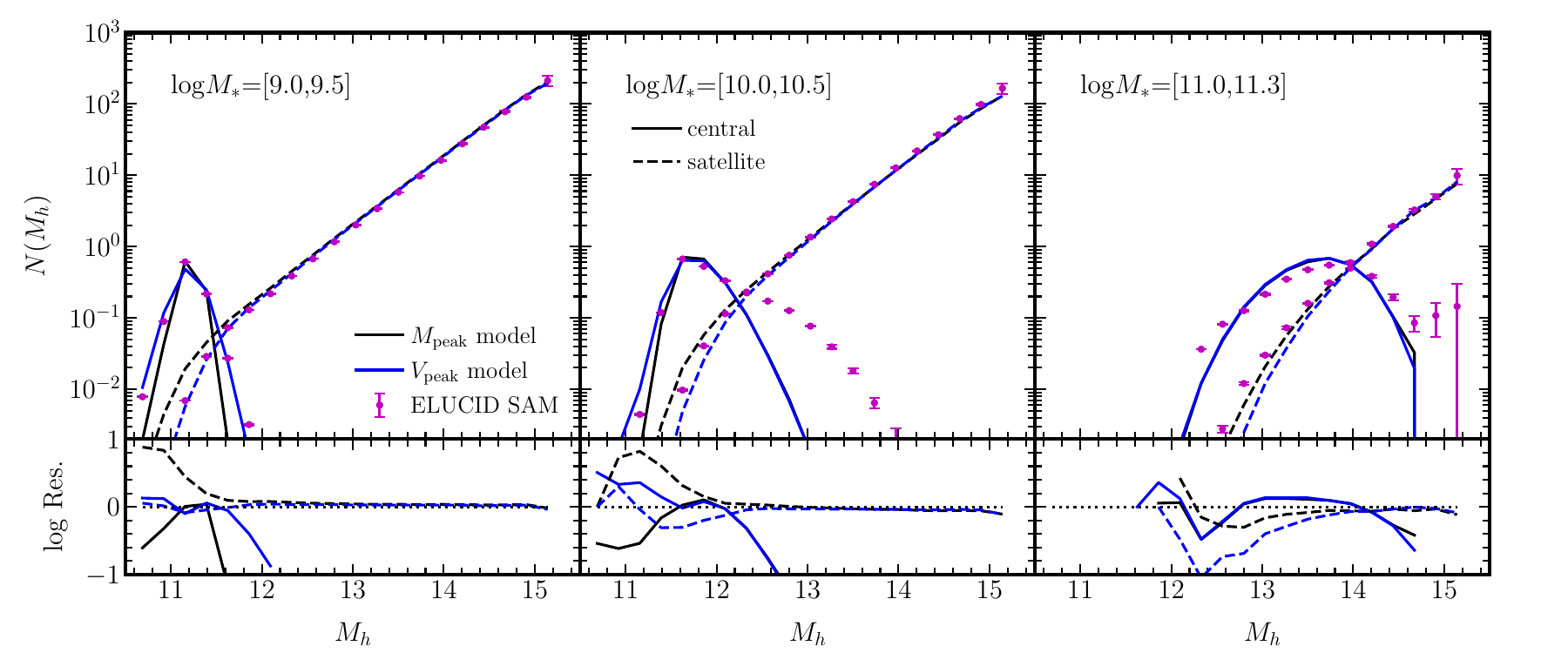}
\includegraphics[width=0.85\textwidth]{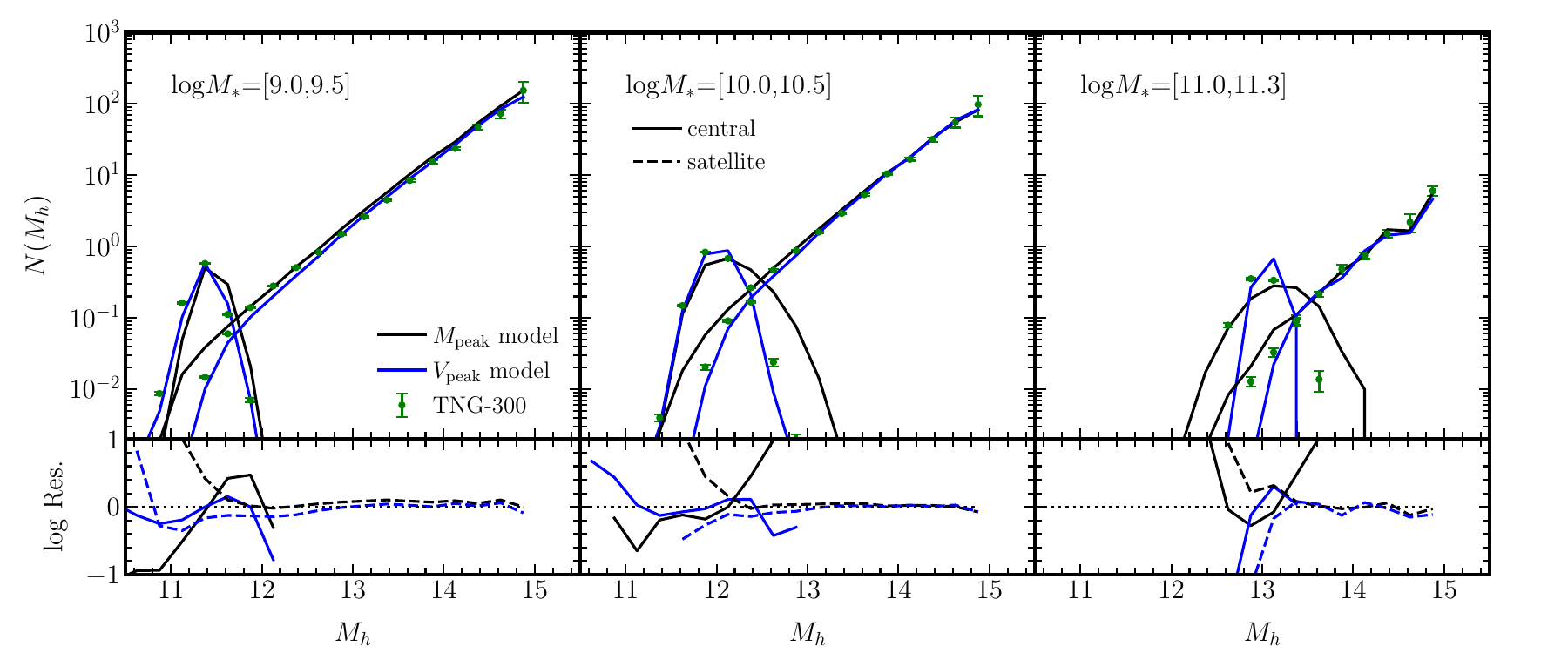}
\caption{Similar to the results shown in Fig. \ref{result_mockhod}, but here for galaxies in different stellar mass bins, ${\rm log}M_*=[9.0,9.5]$, ${\rm log}M_*=[10.0,10.5]$, and ${\rm log}M_*=[11.0,11.3]$ as shown in the left, middle, and right panel, respectively. Results shown in the upper- and lower-panels are for ELUCID SAM and TNG-300, respectively. }
\label{sam_hod}
\end{figure*}

Given that our CS-SHAM framework has already been demonstrated to exert strong constraints on the galaxy distribution—especially on the satellite-galaxy HOD in massive halos—it is natural to ask how stringently it also constrains the CLFs. To explore this, Figure \ref{result_mockclf} presents the CLFs predicted by the best-fit models, which we compare to the “true’’ CLFs measured from the Jiutian mock catalogs in four halo mass intervals, as labeled in the top-left corner of each panel. To compute the CLF, we convert to $z$-band luminosity via ${\rm log}L = 0.4 \times (4.5 - M_z)$. 

Consistent with the HOD analysis, the model that uses the same mass proxy as the Jiutian mock recovers the true CLF very well for both central and satellite galaxies. When alternative subhalo mass indicators are used to constrain the CS-SHAM model, we find modest deviations in $\sigma_L$ from the true values for central galaxies. For satellites, however, the CLF predictions in halos with mass $>10^{13.0}\hmsun$ are remarkably accurate. The only significant discrepancy appears in the lowest halo mass bin, where the model yields a CLF that noticeably differs from the true one. This mismatch can once again be traced back to the fact that the halo bias becomes nearly constant for halo masses below $10^{13.0}\hmsun$, thus the clustering of galaxies has much reduced constraining power.

Overall, regardless of which subhalo mass proxies are adopted for the abundance matching, our CS-SHAM reproduces the true HOD and CLF over a broad range of galaxy luminosities and host halo masses with only four model parameters, providing valuable constraints on the galaxy–halo connection.

\begin{figure*}[!ht]
\centering
\includegraphics[width=0.98\textwidth]{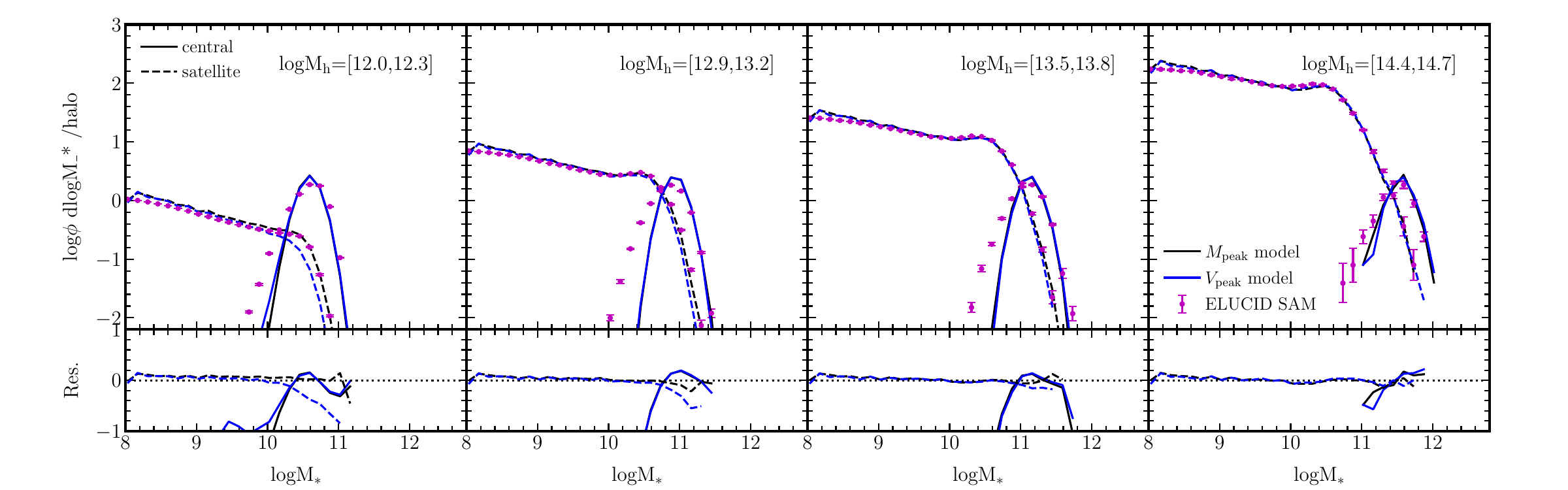}
\includegraphics[width=0.98\textwidth]{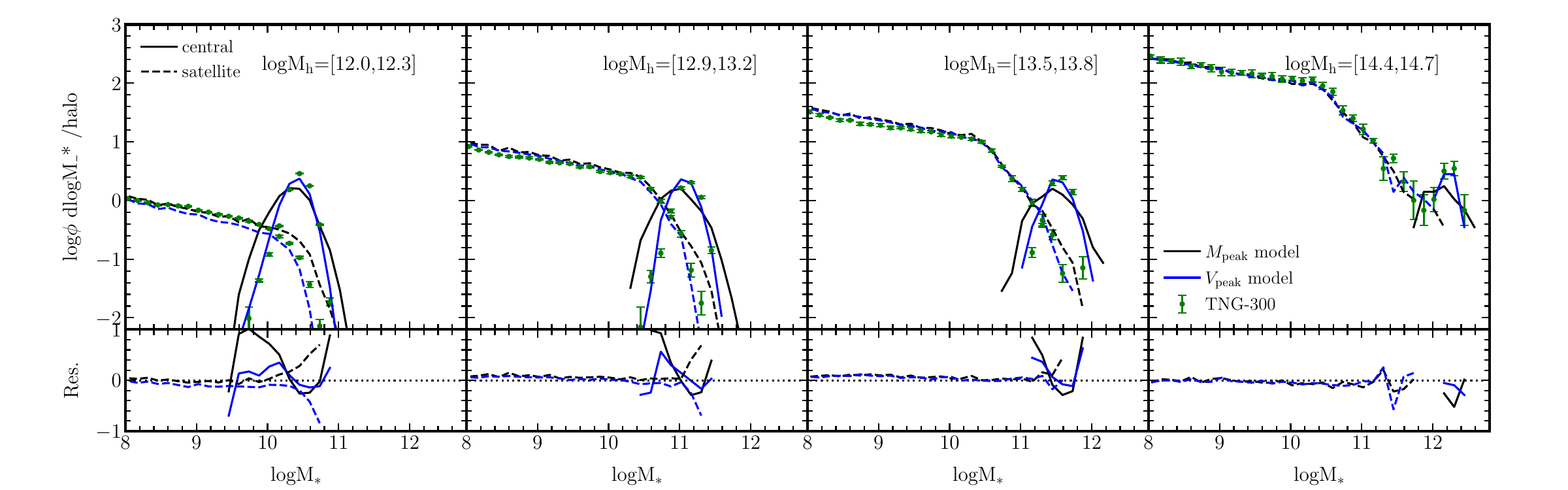}
\caption{Similar to the results shown in Fig. \ref{result_mockhod}, but here for conditional stellar mass functions in halos of different mass ranges. Results shown in the upper- and lower-panels are for ELUCID SAM and TNG-300, respectively. }
\label{fig:sam_tng_csmf}
\end{figure*}

\subsection{HOD and CSMF for ELUCID SAM and TNG-300}
\label{sec:HOD_SAM}

By applying the aforementioned tests to the Jiutian mock catalogs, we demonstrate that CS-SHAM can successfully recover the HOD and CLF of these mocks—originally generated using a basic SHAM approach—through fitting the projected two-point correlation functions across multiple luminosity bins. In this subsection, we further assess how well the CS-SHAM model can constrain the HOD and CSMF for the ELUCID SAM and TNG-300 galaxy samples.

We begin with the ELUCID SAM. Using the best-fit parameters of the $M_{\rm peak}$ and $V_{\rm peak}$ models, we predict the HOD of central and satellite galaxies in three stellar mass bins, as indicated by the labels in the top-left corner of each panel, and compare these predictions with the corresponding true values shown in the upper panels of Figure \ref{sam_hod}.  

For central galaxies, both the $M_{\rm peak}$ and $V_{\rm peak}$ models recover the correct peak halo mass and yield a reasonably accurate scatter in both the low-mass and high-mass bins. However, for the intermediate stellar mass bin shown in the middle panel, the ELUCID SAM central deviates from Gaussian distribution, exhibiting an extraordinary wide tail towards higher host halo masses. This feature, likely arising from detailed galaxy formation physics within the SAM, is not recovered by the $M_{\rm peak}$ or $V_{\rm peak}$ model.

For satellite galaxies, in the lowest stellar mass bin, the $V_{\rm peak}$ model accurately reproduces the SAM HOD. The $M_{\rm peak}$ model predicts  slightly  more satellite galaxies than the SAM below ${\rm log}M_h \sim 12$. In the intermediate stellar mass bin,  the satellite numbers are reproduced by both models above $\log M_h \sim 12.5$, with minor discrepancies appearing at lower masses. For the most massive stellar bin, both models accurately match the satellite occupation for halos with ${\rm log}M_h>14$, yielding $\log \rm Res. \sim 0.04$.

Similar to the procedure used for the ELUCID SAM, we predict the HODs of central and satellite galaxies using the best-fit parameters and compare them with those derived from TNG-300 in the lower panels of Figure \ref{sam_hod}. The overall performance is very similar to that of the ELUCID SAM. Overall, the $V_{\rm peak}$ model provides a slightly better fit to the TNG-300 HODs than the $M_{\rm peak}$ model, achieving satellite occupation predictions with an accuracy of $\log \rm Res. \sim 0.04$ for halos with $\log M_h \gtrsim 13$.

We also present the predicted conditional stellar mass functions (CSMFs) in four halo mass bins obtained using the best-fit parameters for the ELUCID SAM in the upper panels of Figure \ref{fig:sam_tng_csmf}. The $\mpeak$ and $\vpeak$ models yield very similar central CSMFs across all halo mass bins, and both are slightly narrower than those from the ELUCID SAM. For satellite galaxies, both models successfully reproduce the ELUCID SAM CSMF in halos with $\log M_h \gtrsim 12.9$. Even in the lowest halo mass bin, the discrepancy between the models and the ELUCID SAM remains small.

The lower panels of Figure \ref{fig:sam_tng_csmf} show the predicted CSMFs for TNG-300. The overall behavior is very similar to that found for the ELUCID SAM: we achieve highly accurate CSMF predictions for halos with masses above $\log M_h \sim 12.9$, with only a small discrepancy appearing in the lowest halo mass bin. Overall, the $\vpeak$ model yields a slightly better agreement with the central CSMF.


\section{Summary and discussion}
\label{sec:summary}

In this study, we introduce a CS-SHAM framework that treats central and satellite galaxies independently, characterized by a satellite fraction that depends on stellar mass or magnitude. The satellite fraction, $f_{\rm sat}$, is described by an error function with three free parameters, and the model further includes a scatter parameter in the relation between stellar mass and subhalo mass. We test our model using four sets of mock catalogs: two Jiutian mocks generated with basic SHAM based on the subhalo mass proxies $M_{\rm peak}$ and $V_{\rm peak}$, one ELUCID SAM mock, and one TNG-300 mock. For each mock catalog, we apply CS-SHAM with the two subhalo mass indicators to fit the projected two-point correlation function $w_p$ and thereby constrain the model parameters.

We find that, for a mock built using a particular mass indicator, fitting galaxy clustering in different luminosity and stellar mass bins with a model based on that same indicator can accurately recover $f_{\rm sat}$ and, in turn, the halo occupation distribution and conditional luminosity function. In contrast, models that rely on alternative mass indicators do not always recover the true $f_{\rm sat}$. Nonetheless, even when $f_{\rm sat}$ is not perfectly recovered, the HOD and the CLF/CSMF of galaxies can still be tightly constrained, especially in massive halos.  By combining the results from the Jiutian SHAM mocks, the ELUCID SAM, and TNG-300, we demonstrate that within our CS-SHAM framework with four free parameters—regardless of whether it is implemented with an $M_{\rm peak}$ or a $V_{\rm peak}$ model—galaxy clustering alone can faithfully reproduce both the CLFs and CSMFs in halos more massive than $\log M_h \sim 12.9$.  

To further demonstrate the importance of explicitly incorporating $f_{\rm sat}$ modeling in our new CS-SHAM framework, we carry out both traditional SHAM and CS-SHAM with a true $f_{\rm sat}$ curve to predict $w_p$ for ELUCID SAM and TNG-300. The corresponding results are presented in Appendix \ref{sec:sigmaonly} and \ref{sec:trueparam}, respectively. It is clear that traditional SHAM fails to accurately recover $w_p$, particularly when the $M_{\rm peak}$ model is adopted. Moreover, even within our new CS-SHAM framework, when the true $f_{\rm sat}$ curve is used, the predictions remain sensitive to the choice of mass tracer. Except for the special case already shown (e.g., an $M_{\rm peak}$ SHAM mock analyzed with the $M_{\rm peak}$ model), neither $M_{\rm peak}$ nor $V_{\rm peak}$ alone can achieve a fit quality comparable to the case where $f_{\rm sat}$ modeling is incorporated.

Given the performance of CS-SHAM, we plan to apply this model to observational samples such as DESI and future CSST data. When these clustering constraints are further combined with group and cluster observables, including weak lensing, X-ray, and SZ measurements, they will allow stringent tests of cosmological models. Moreover, this model can be deployed in large-volume cosmological simulations to build mock catalogs that simultaneously match the observed galaxy luminosity function and clustering at low computational cost, which is particularly valuable for constructing emulators for cosmological analyses.

Beyond these current applications, our model can be generalized in several ways to gain additional flexibility and enable wider use cases. First, regarding parameterization, we can introduce an extra parameter to describe the scatter of the SHMR as a function of halo mass, as well as another parameter to characterize $f_{\rm sat}$ at the faint end, thereby allowing the generation of dwarf galaxies in mock catalogs used to study galaxy formation. Next, considering the established correlation between galaxy properties and halo properties beyond mass in SAMs and hydrodynamic simulations, we can introduce a parameter to account for galaxy assembly bias. This would allow for a more comprehensive investigation of the galaxy-halo connection. Furthermore, similar to other modified SHAM approaches in the literature that include orphan galaxies to improve small-scale clustering predictions, orphan galaxies can also be incorporated within our framework. Because disrupted subhalos that host orphan galaxies may exhibit radial distributions within halos that differ from those of surviving subhalos, including them has the potential to further refine the modeling of $w_p$.

In this work, we concentrate on the full galaxy population. Observationally, however—particularly in cosmological applications—only a subset of galaxies is actually targeted, such as ELGs or LRGs, which requires an additional layer of modeling. For instance, \cite{Yu2024} introduce a generalized SHAM framework that treats $f_{\rm sat}$ as a free parameter to describe ELG clustering in the DESI One Percent Survey. They demonstrate that $f_{\rm sat}$ plays a key role in reproducing the quadrupole of DESI ELGs. Similarly, Favole et al. (2025) employ an $f_{\rm sat}$ parameter to model the auto- and cross-correlation functions of ELGs and LRGs, enabling them to constrain the satellite fraction of ELG/LRG galaxies around centrals of different types. \cite{Gao2023} model the auto- and cross-correlation of DESI ELGs using an SHMR-based approach with parameters that encode the probability that a halo of a given mass hosts an ELG. By introducing extra parameters to capture color or type segregation, our CS-SHAM framework can be straightforwardly generalized to describe the stellar-mass-dependent clustering of ELGs and LRGs.

In addition, analogous to standard SHAM frameworks, our model can be generalized to conditional abundance matching (CAM; \cite{Hearin2013,Watson2015,Contreras2021b,DeRose2022,Favole2022}), in which secondary galaxy properties such as color are linked to secondary halo properties such as formation time. This allows one to predict how galaxy clustering varies with color or star formation rate (SFR). One can likewise associate emission-line luminosities with halo properties in order to reproduce the joint dependence of ELG clustering on both stellar mass and emission-line luminosity \cite{Hagen2025}. These extensions will not only yield new insights into the galaxy–halo connection and the physics of galaxy formation, but will also lessen the need for additional parameterization for ELGs and LRGs, thereby increasing the effectiveness of CS-SHAM model for cosmological studies when used alongside large-volume simulation suites.

\Acknowledgements{This work is supported by the National Key R\&D Program of China (2023YFA1607800, 2023YFA1607804, 2023YFA1605600, 2023YFA1607801), the National Science Foundation of China (No. 12373003, 12595312), “the Fundamental Research Funds for the Central Universities”, 111 project No. B20019, and Shanghai Natural Science Foundation, grant No.19ZR1466800. XX acknowledge the support of National Science Foundation of China (no.12503007). We acknowledge the science research grants from the China Manned Space Project with Nos. CMS-CSST-2021-A02, CMS-CSST-2021-A03 \& CMS-CSST-2025-A04, and Yangyang Development Fund. This project is also supported in part by Office of Science and Technology, Shanghai Municipal Government (grant Nos. 24DX1400100, ZJ2023-ZD-001).}

\InterestConflict{The authors declare that they have no conflict of interest.}




\begin{appendix}
\section{}
\renewcommand{\thesection}{Appendix}

\subsection{\label{sec:sigmaonly} Clustering predictions of traditional SHAM}

\begin{figure*}[!ht]
\centering
\includegraphics[width=0.85\textwidth]{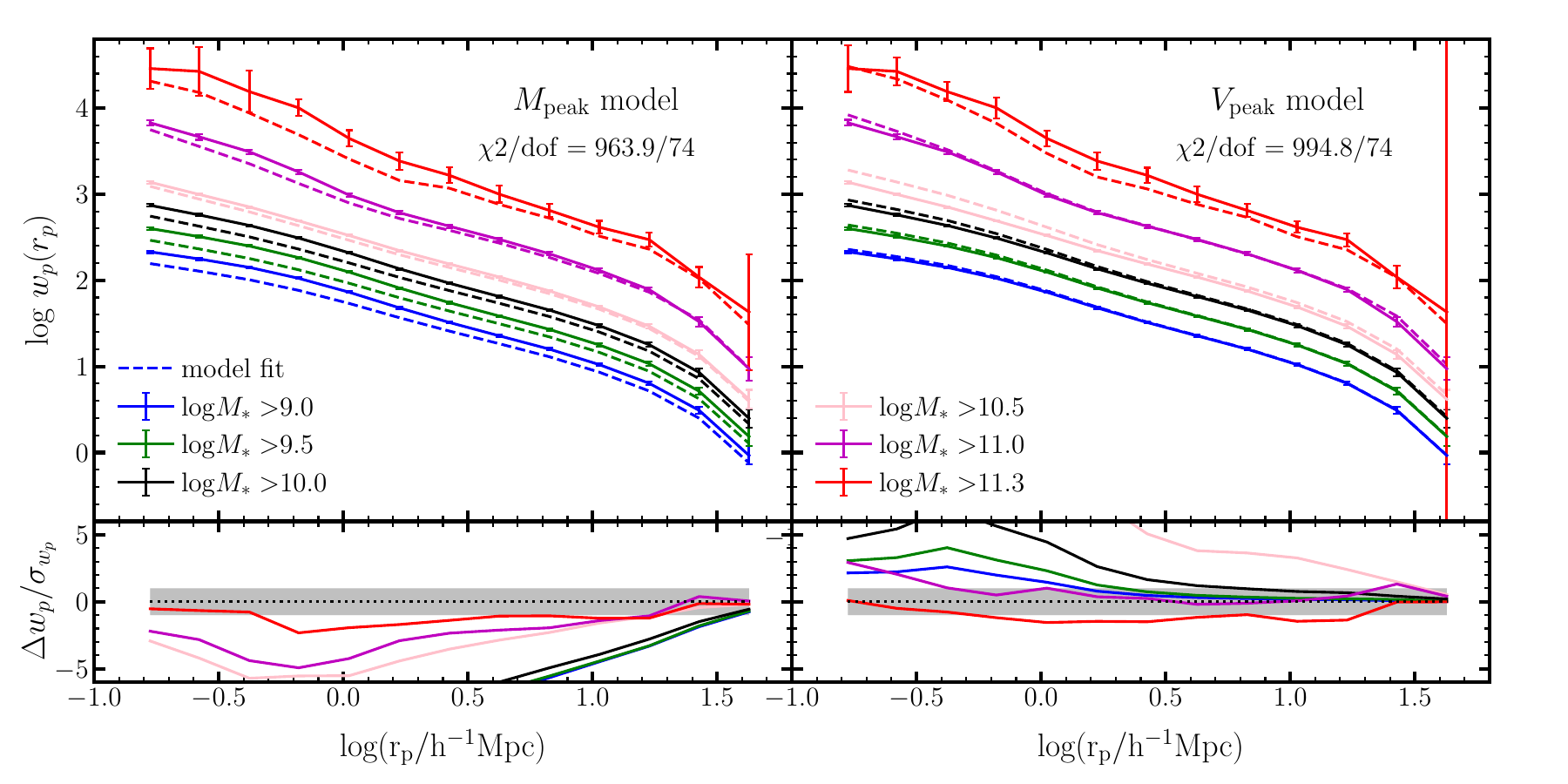}
\includegraphics[width=0.85\textwidth]{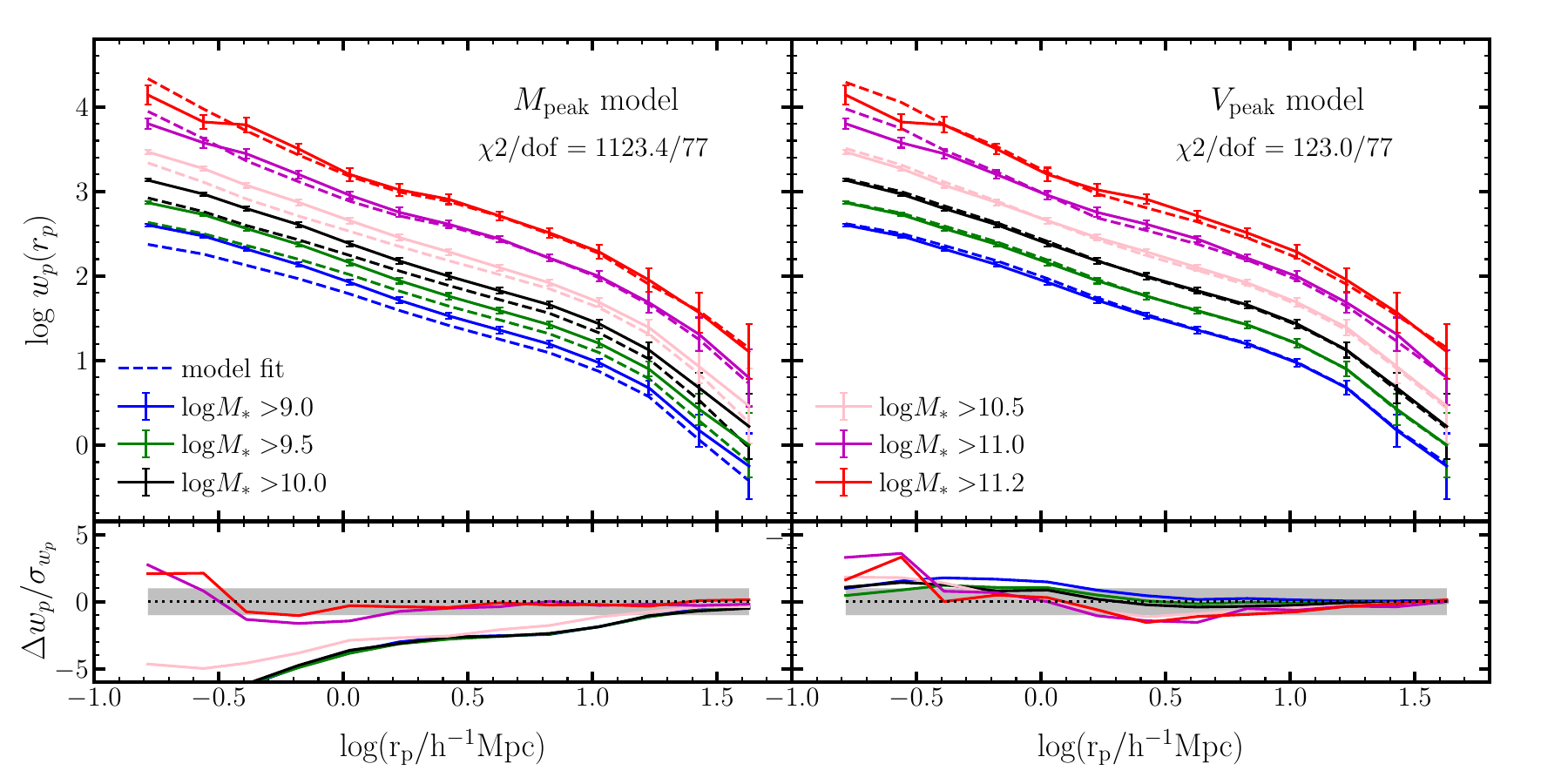}
\caption{$w_p$ prediction using the basic SHAM with only the $\sigma$ parameter for the ELUCID SAM (top) and TNG-300 (bottom) with $M_{\rm peak}$ model (left) and the $V_{\rm peak}$ model (right).}
\label{sam_wp_sigonly}
\end{figure*}

In this appendix, we show the results of fitting $w_p$ of ELUCID SAM and TNG-300 using a basic SHAM model that incorporates only the $\sigma$ parameter in Figure \ref{sam_wp_sigonly}. For the ELUCID SAM, the $M_{\rm peak}$ model underpredicts $w_p$ for all stellar mass threshold bins. For the two lowest mass bins, these deviations are larger than $6\sigma_{w_p}$. In contrast, the $V_{\rm peak}$ model overpredicts $w_p$ except for the most massive bin. For TNG-300, the $M_{\rm peak}$ model again underpredicts $w_p$ on most scales, with particularly large deviations ($>6\sigma_{w_p}$) for low-mass bins. The $V_{\rm peak}$ model provides a more reasonable prediction achieving $\chi^2/{\rm dof}=1.6$, and most of the deviations are within $1\sigma_{w_p}$. The significant differences in the performance of the basic SHAM model reflect its limited flexibility. With a single $\sigma$ parameter which primarily affects the clustering of massive galaxies, the model offers little control over the clustering of low-mass galaxies. The CS-SHAM addresses this issue by specifying the fraction of satellite galaxies.

\subsection{\label{sec:trueparam} Clustering predictions using fixed true $f_{\rm sat}$ in CS-SHAM}

\begin{figure*}[!ht]
\centering
\includegraphics[width=0.85\textwidth]{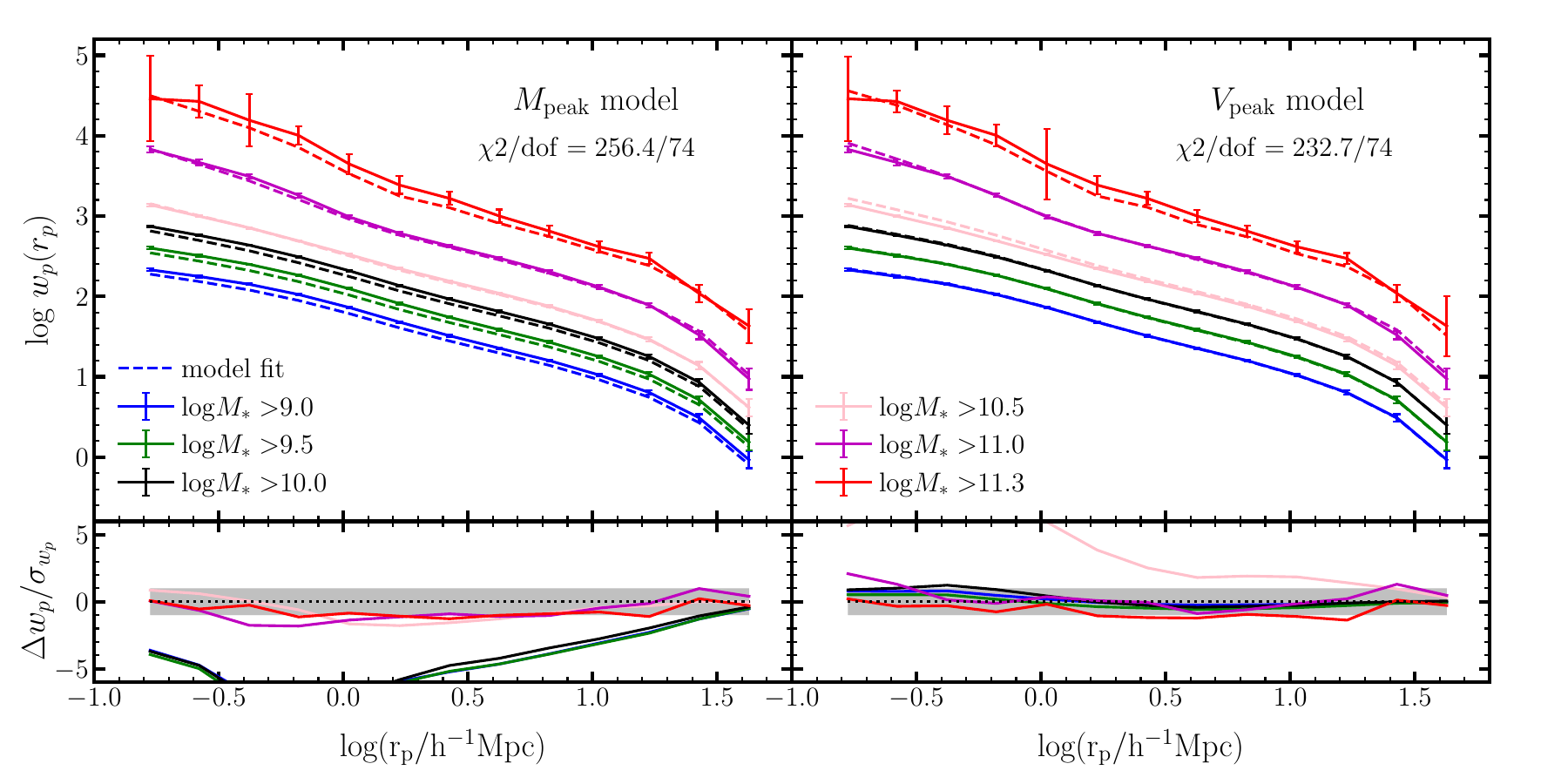}
\includegraphics[width=0.85\textwidth]{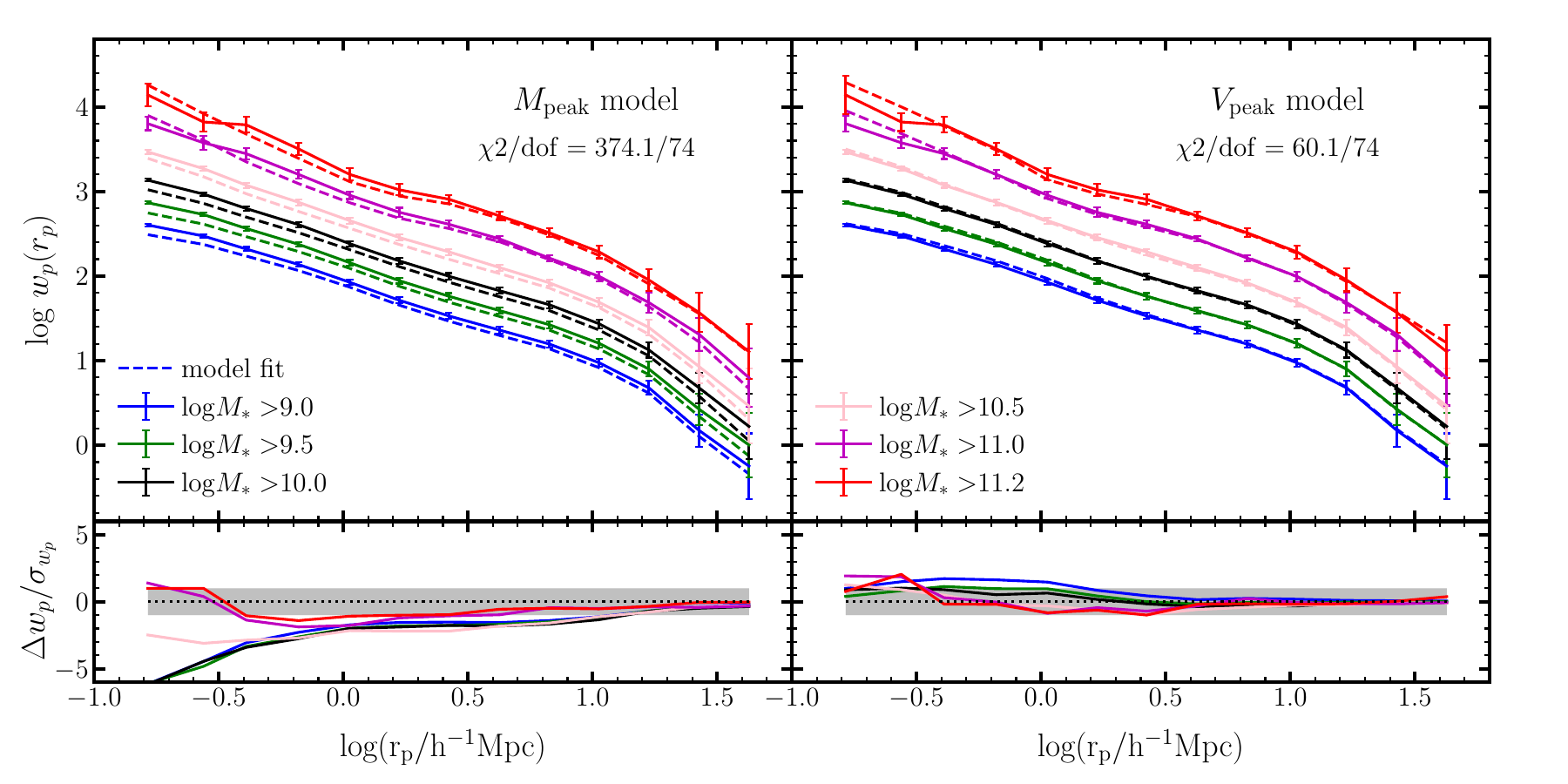}
\caption{$w_p$ prediction using the underlying true CS-SHAM parameters of the ELUCID SAM (top) and TNG-300 (bottom) with $M_{\rm peak}$ model (left) and the $V_{\rm peak}$ model (right).}
\label{sam_wp_trueparam}
\end{figure*}

In Figure \ref{sam_wp_trueparam}, we present the $w_p$ predictions obtained by applying the true $f_{\rm sat}$ curve within the new CS-SHAM framework to the ELUCID SAM and TNG-300, for both the $M_{\rm peak}$ and $V_{\rm peak}$ models. For the ELUCID SAM, the $V_{\rm peak}$ model recovers $w_p$ mostly within $1\sigma$, except for the bin of ${\rm log}M_*>10.5$ (which results in a large $\chi^2$). The $M_{\rm peak}$ model underpredicts the $w_p$ for the two faintest bins, though it recovers the $w_p$ for other bins within $2\sigma$. For TNG-300, the performance of the $V_{\rm peak}$ model is similar to that using the best-fit parameters in Figure \ref{sam_wp}, recovering $w_p$ within $2\sigma$ with $\chi^2/{\rm dof}<1$. However, $M_{\rm peak}$ model systematically underpredicts the $w_p$ for all stellar mass bins except at small scales. These findings support the conclusion in Section \ref{sec:fitsam} that the $V_{\rm peak}$ model successfully recovers $w_p$ while yielding a more realistic $f_{\rm sat}$, whereas the $M_{\rm peak}$ model requires a higher $f_{\rm sat}$ to match $w_p$. Overall, based on the $\chi^2/{\rm dof}$ values, we find that, comparing to the traditional SHAM,  implementing the true $f_{\rm sat}$ curve within the CS-SHAM framework provides a better description of galaxy clustering. Nevertheless, these $\chi^2/{\rm dof}$ values remain higher than in the case where $f_{\rm sat}$ is explicitly modelled. We therefore infer that only by pairing a suitable mass indicator with its corresponding true $f_{\rm sat}$ curve can one obtain the tightest constraints on clustering, as well as on the HOD and CLF. In practice, however, neither of these quantities is directly measurable from observations.

\end{appendix}

\end{multicols}

\end{document}